\documentclass[reprint,
superscriptaddress,
amsmath,amssymb,aps,prc,
]{revtex4-2}

\usepackage{graphicx}
\usepackage{dcolumn}
\usepackage{bm}
\usepackage[mathlines]{lineno}
\usepackage{color,xcolor}
\usepackage{tikz, pgfplots}
\usepackage{listofitems} 
\usetikzlibrary{arrows.meta} 
\usepackage[outline]{contour} 
\contourlength{1.4pt}

\tikzset{>=latex} 
\usepackage{xcolor}
\tikzstyle{node}=[thick,circle,draw=black,minimum size=22,
inner sep=0.5,outer sep=0.6]
\tikzstyle{node in}=[node,black,draw=black,fill=none]
\tikzstyle{node hidden}=[node,black,draw=black,fill=none]
\tikzstyle{node out}=[node,black,draw=black,fill=none]
\tikzstyle{connect}=[thick,black] 
\tikzstyle{connect arrow}=[-{Latex[length=4,width=3.5]},thick,black,
shorten <=0.5,shorten >=1]
\tikzset{ 
  node 1/.style={node in},
  node 2/.style={node hidden},
  node 3/.style={node out},
}
\def\nstyle{int(\lay<\Nnodlen?min(2,\lay):3)} 
\newcommand{\EIA}{
Grupo de Física Teórica y Aplicada, EIA University, Envigado, Colombia}
\newcommand{\guane}{
guane enterprises, R+D+I Unit, Medell\'in, Colombia.}
\newcommand{\UIS}{
Universidad Industrial de Santander, 
Grupo de O\'ptica y Tratamiento de Señales, Bucaramanga, Colombia.}
\newcommand{\SCU}{
Sivas Cumhuriyet University, Faculty of Science, Department of Physics,
Sivas, 58140, Turkey.
}
\newcommand{\GFTyMA}{
Grupo de F{\'i}sica Te{\'o}rica y Matem{\'a}tica Aplicada, Instituto de F{\'i}sica, 
Facultad de Ciencias Exactas y Naturales, Universidad de Antioquia, Medell\'in, Colombia.}

\begin{document}

\title{Predicting $\beta$-Decay Energy with \\Machine Learning}

\author{Jose M. Munoz}
\email{jose.munoz25@eia.edu.co}
\affiliation{\EIA}
\affiliation{\guane}

\author{Serkan Akkoyun}
\email{sakkoyun@cumhuriyet.edu.tr}
\affiliation{\SCU}

\author{Zayda P. Reyes}
\email{zaydaprq@correo.uis.edu.co}
\affiliation{\guane}
\affiliation{\UIS}

\author{Leonardo A. Pachon}
\email{leonardo.pachon@udea.edu.co}
\affiliation{\guane}
\affiliation{\GFTyMA}

\date{\today}
\begin{abstract}
$Q_\beta$ represents one of the most important factors characterizing 
unstable nuclei, as it can lead to a better understanding of nuclei 
behavior and the origin of heavy atoms. 
Recently, machine learning methods have been shown to be a powerful tool to
increase accuracy in the prediction of diverse atomic properties such as 
energies, atomic charges, volumes, among others.
Nonetheless, these methods are often used as a black box not 
allowing unraveling insights into the phenomena under analysis.
Here, the state-of-the-art precision of the $\beta$-decay energy on 
experimental data is outperformed by means of an ensemble of machine-learning models.
The explainability tools implemented to eliminate the black box concern 
allowed to identify uncertainty and atomic number as the
most relevant characteristics to predict $Q_\beta$ energies. Furthermore, physics-informed feature
addition improved models' robustness and raised vital characteristics of theoretical models 
of the nuclear structure.
\end{abstract}

\maketitle

\section{\label{sec:Intro} Introduction}
There are about 250 stable isotopes and over 3000 unstable ones are
known. 
%
%
One way that unstable atomic nuclei decay into stable ones is $\beta$-
decay.
As a result, the $\beta$-decay energy $Q_\beta$ that goes with it is a 
fundamental property of unstable atomic nuclei.
$Q_\beta$ values can be determined via several methods, such as 
$\beta$ endpoint measurements, counting in coincidence with annihilation 
radiation, electron capture $EC/\beta^+$ ratio method, $\gamma$ 
absorption with X-ray coincidence \cite{alkhazov1993beta}. 
This process is complex to explore provided that its energy spectrum has a
continuous structure.
The decay energy seems to simply relate to the proton number and mass of
the atomic nuclei. 
Experimental $Q_\beta$ energies of nuclei can be verified in the 
$Z \in [4,\ 82]$, $N \in [4,\ 126]$ regions, whereas unknown energies can 
be calculated using $\alpha$-$\beta$ energy cycles by means of reliable 
$\alpha$ spectroscopic data. 
Moreover, some $\beta$-decay energies can be obtained with the help of 
energy cycles from $\alpha$-decay energy systematic 
\cite{kolesnikov1959proton}. 

\subsection{Explainable Machine Learning}
\label{subsec:Explainable-ML}
Machine Learning (ML) has recently emerged as an important tool for 
understanding physical phenomena, focusing on understanding how a model 
makes a particular decision, rather than just accepting the output as a 
black-box \cite{Carleo:2019ptp, karniadakis2021physics}. 
In the supervised approach, an algorithm is used as a functional 
approximation $\mathcal{F}(x)$ for an unknown function 
$F(x)$. 
Thus, it is possible to obtain the desired output for the collected 
physical data $x\in \mathbb{R}^n$.
Classical statistical algorithms such as linear regression models or 
Principal Components Regression (PCR) allow for characterizing the 
relevant predicting features of the model, thus providing an understanding
of the physical phenomena underlying the data. 
However, they impose specific restrictions on the studied manifold, and 
thus, they are not well suited to most of the complex phenomena observed in 
nature \cite{hastie2009elements}.
On the other hand, modern approaches such as deep learning (DL) allow for
a comprehensive representation of physical observables, as they work 
as universal function approximators \cite{chen2018universal, 
scarselli1998universal}. 
The trade-off comes when it is necessary to understand what the algorithms
are considering. 
To address this issue, explainable ML allows identifying of underlying 
characteristics such as feature importance 
\cite{casalicchio2018visualizing, 
wojtas2020feature,rengasamy2021mechanistic}, model the dynamics of complex 
systems \cite{cranmer2020lagrangian, chen2018neural}, conservation law 
discovery \cite{liu2021machine, lee2019deep} and symbolic expression 
extraction \cite{cranmer2020discovering, udrescu2020ai, 
bakarji2022discovering}.
Besides, simple ML models have previously been used in the study of the 
$\beta$-decay~\cite{niu2019predictions, bayram, cruz2019clustering, 
costiris2007global, gao2021machine}; however, there is still the need to 
obtain physical intuition out of these techniques. 
Nuclear physics is not an exception, as determination of one and two 
proton  separation energies \cite{athanosso-1}, developing nuclear mass 
systematics  \cite{athanosso-2}, determination of ground state energies of 
the nuclei \cite{bayram}, identification of impact parameters in heavy-ion 
collisions \cite{Bass:1996ez}, estimation of fusion reaction cross-section 
\cite{nimb}, estimating nuclear RMS charge radii \cite{radii-1}, decoding 
$\beta$-decay systematic \cite{cost} and determination of fusion barrier 
heights \cite{barrier} have already been studied via this tooling.

\subsection{ML for $\beta$-decay}
\label{subsec:ML_beta_decay}
To be precise, the $\beta$-decay energies \cite{beta} were estimated by 
considering the neutron and proton numbers of atomic nuclei as the only 
input parameters and the data values have been obtained based on the 
Hartree–Fock–BCS method with the Skyrme force MSk7 \cite{Tondeur}.
By contrast, here, the recent AME2020 experimental data \cite{AME2020} is 
considered and the input parameters are augmented by considering a full 
range of physical properties of atoms (see Table~\ref{tab:table_desc} 
below).
By using the experimental $Q_\beta$ values available in the literature, 
the systematics of atomic nuclei related to this energy was obtained here 
by performing ML, considerably decreasing the root mean 
squared error (RMSE) of previous approaches by a factor of 2.5.
In addition, by calculating the importance scores, the fundamental 
parameters of atomic nuclei that have an impact on the $Q_\beta$ energies 
are highlighted. 
%

\subsection{The atomic mass table}
\label{subsec:Intro_A}
%
The dataset consists of the classical nuclear variables, numerically 
described in Table~\ref{tab:table_desc}, whence entries with any 
non-computable data entrances were removed, leaving a total of 2813 
unstable isotopes. 
The relevant dataset variables are the number of neutrons in each isotope
$N$, number of protons in each isotope $Z$, mass number $A$, difference 
between measured mass and the mass number $M_e$ and binding energy $B_E$.
Here, notice that a majority of the features are highly skewed. 
Further description of the raw data is shown in 
Appendix~\ref{sec:datadist}.
\begin{table*}
\begin{tabular}{|lrrrrrrrrrrr|}
\hline \hline
{} &           $N-Z$ &            $N$ &            $Z$ &           $m$ 
Excess &     $\delta_m$ &      Binding E &          $Q_\beta$ &     
$\delta_Q\beta$ &            $p$ &    Atomic $m$ &       $\delta_am$ 
\\
\hline
mean  &    27.783 &    86.782 &    58.999 &  -24262.379 &   115.570 &  
7985.498 &   1546.938 &   149.663 &   145.043 &  713121.060 &   120.919114 
\\
std   &    16.501 &    42.272 &    27.627 &   59122.015 &   239.781 &   
666.922 &   7441.595 &   231.639 &    69.683 &  388483.889 &   199.856141 
\\
min   &    -3 &     1 &     0 & -341875.725 &     0.000 &     0.000 & 
-16673.000 &     0.000 &     1.000 &      99.560 &     0.000 
\\
25\%   &    14 &    53 &    38 &  -67307.280 &     2.251 &  7731.891 &  
-4103.108 &     5.004 &    91.000 &  171690.750 &     2.417 
\\
50\%   &    26 &    85 &    59 &  -40851.305 &    10.614 &  8075.771 &     
62.237 &    22.526 &   144.000 &  926617.954 &    11.394 
\\
75\%   &    40 &   119 &    81 &    2679.998 &   179.083 &  8383.611 &   
6087.989 &   206.267 &   199.000 &  954880.438 &   189.704 
\\
max   &    64 &   177 &   117 &  196397.000 &  8013.457 &  8794.555 &  
32740.000 &  1082.000 &   294.000 &  999981.252 &  1078.000 
\\
\hline \hline
\end{tabular}
\caption{Numerical distribution of the AME2020 data for each one of the 
measured parameters.}
\label{tab:table_desc}
\end{table*}

It is possible to identify high correlations between some of the variables 
provided that some of the features are computable from the others 
(such as the mass number from the atomic and nuclear numbers). 
This needs to be addressed since it is known that algorithms that are 
susceptible to multicollinear features generate unreliable predictions
\cite{garg2012comparison}. 

Therefore, an exploratory dimensionality reduction analysis is implemented
using Principal Component Analysis (PCA) with the normalized variables
and evaluating the performance of a naive predictor model. 
Here, both the explained variance and a principal component 
regression (PCR) used a 5-fold cross-validation scheme 
\cite{jolliffe1982note}. 
As expected, more than two principal components are necessary to
give a complete sense of $Q_\beta$ by evaluating the RMSE of a linear 
regression model. 
In addition, Fig.~\ref{fig:r2pca} depicts the asymptotic behavior that is 
reached using more than 3 principal components. 
Nevertheless, neither the second nor the third principal components are 
negligible.

\begin{figure}[!htb]
\label{fig:r2pca}
   \includegraphics[width=0.4\textwidth]
   {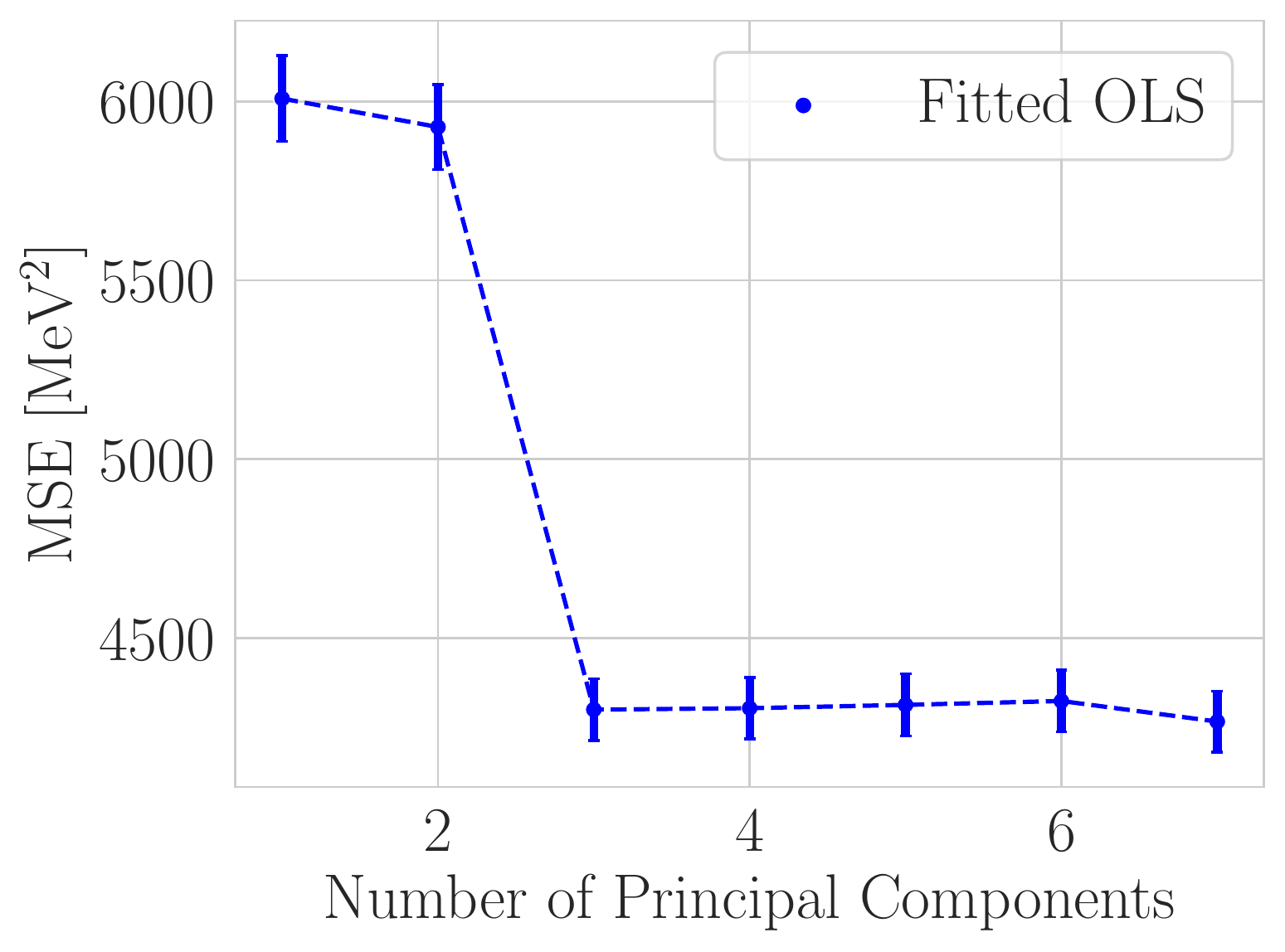}
   \caption{\centering RMSE of the fit for a different number of components
   in the PCA analysis.}
   \label{fig:r2pca}
\end{figure}

\begin{figure}[!h]
   \centering
   \includegraphics[width=0.5\textwidth]{
   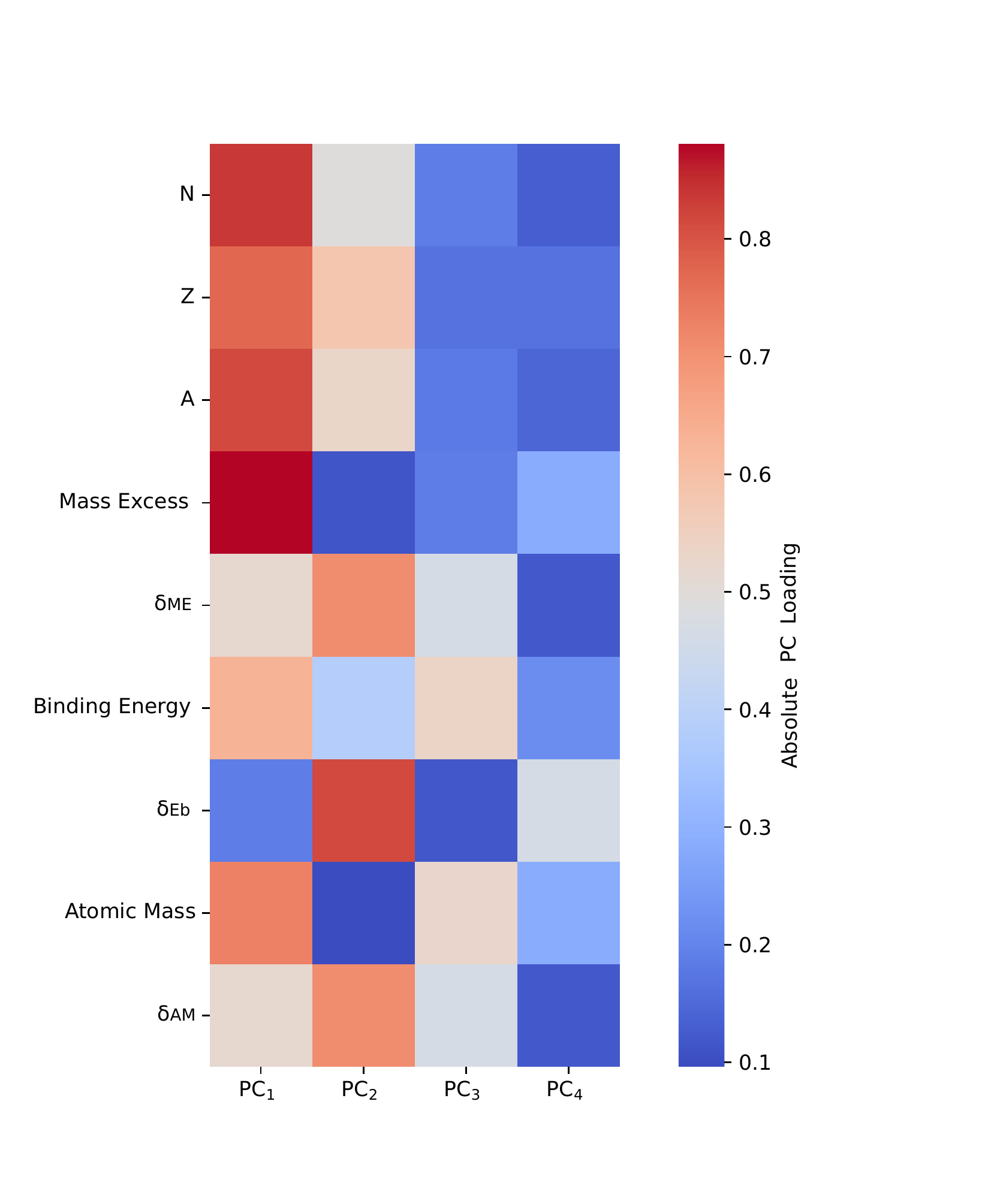}
   \caption{\centering Loadings of each one of the variables into the 
   principal orthogonal  components. See text for more details.}
   \label{fig:loadings_pca}
\end{figure}

\begin{figure}[!h]
   \centering
   \includegraphics[width=0.5\textwidth]{
   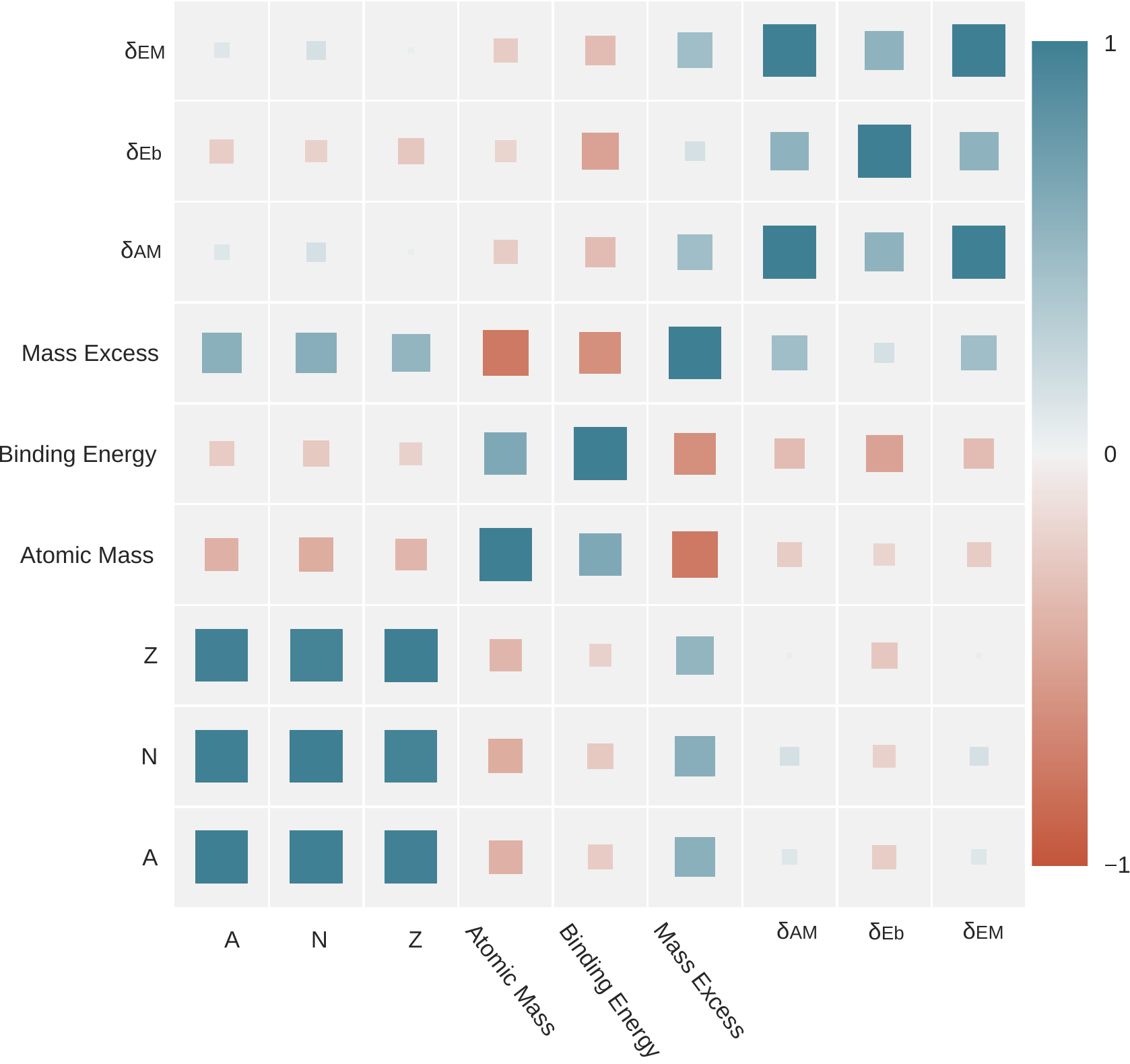}
   \caption{Correlations between variables in the Atomic Mass table.}
   \label{fig:correlations}
\end{figure}

The PCA loadings, obtained as shown in Fig.~\ref{fig:loadings_pca}, were 
used to understand the importance of each of the parameters.
Interestingly enough, the second orthogonal component loads more heavily 
the uncertainties, rather than the components themselves, with a major
role in the binding energy $B_\mathrm{E}$. 
This suggests that there exists a significant contribution of the features 
related to the experimental uncertainties reported in the original dataset. 
However, the correlation analysis shown in Fig.~\ref{fig:correlations} 
accounts for part of this effect. 
Nonetheless, it is expected that Universal Approximators can easily 
overcome this issue.

\section{Modeling $Q_\beta$: A regression approach}
To start the modeling of the phenomena in a systematic and replicable way, 
training/testing/validation sets were generated with the dataset using a 
70/20/10 split. 
A variety of tabular models were fitted in an automatic manner, 
reporting the scores, and using only the training and validation splits. 
In this approach, in each one of the steps, the best-performing models are 
selected and their hyper-parameters are tuned via Bayesian optimization 
techniques as designed by Ref.~\cite{DBLP:journals/corr/abs-1907-10902}. 
%
%
Based on a collection of regression metrics (such as explained variance,
RSME, and their volatility among different splits of the training set), the 
models obtaining high scores while keeping explainability were selected and 
further explored via hyperparameter tuning and feature importance 
measurement techniques.

To have a baseline, the previous work \cite{beta} was replicated using a 
similar approach. 
Here, a vanilla fully-connected Multi-Layer Perceptron (MLP) was 
implemented, but with the AME2020 dataset, which was trained only using 
the $N$ and $Z$ of each one of the entries. 
The architecture consists of a single hidden layer with 100 neurons, 
hyperbolic-tangent non-linearities, and classic Stochastic Gradient 
Descent (SGD).

\subsection{ML and AI models}
\label{subsec:ML_AI_models}
\subsubsection{Ordinary Linear Regression}
\label{subsec:ordinary_regression}
Besides its simplicity, this approach is well-suited to scientific tasks
because of the transparency, a general regression allows. 
Thus, it serves both as a benchmark and brings insights into the process. 
The Ordinary Least Squares (OLS) approach uses a projection model to fit 
the input vector $\mathbf{X}$ containing the selected features to obtain an 
estimator $\mathcal{Q}_\beta$ via
\begin{equation}
\label{equ:q_beta_fit}
\mathcal Q_\beta = \mathbf{X}^T \cdot \mathbf{k} + \varepsilon.
\end{equation}
The optimal parameters for the vector of coefficients $\mathbf{k}$ and 
the scalar $\varepsilon$, which minimizes the difference between the 
observation $Q_\beta$ and the outputs $\mathcal Q_\beta$ are obtained via 
the Ordinary Least Squares procedure.

Moreover, this technique allows for obtaining a statistical measurement 
of the fitted coefficients and their uncertainties, thus, getting insight 
into the relevant features and characteristics of the model.

\subsubsection{Modern MLP}
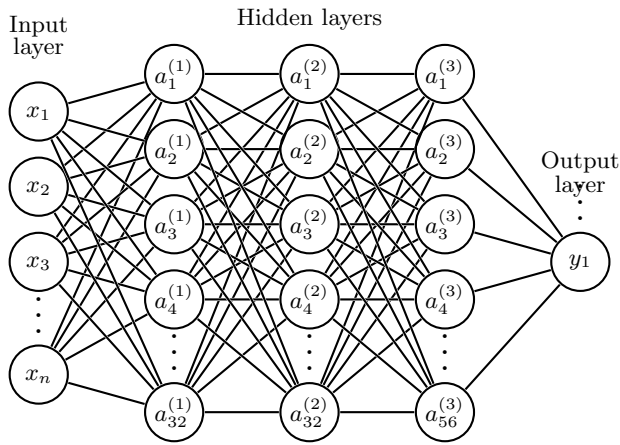
\begin{figure}[h]
\centering
\begin{tikzpicture}[x=1.8cm,y=1cm][transform canvas={scale=0.6}]
  \message{^^JNeural network, shifted}
  \readlist\Nnod{4,5,5,5,1} 
  \readlist\Nstr{n,32,32,56,1} 
  \readlist\Cstr{\strut x,a^{(\prev)},a^{(\prev)},a^{(\prev)},y} 
  \def\yshift{0.5} 
  
  \message{^^J  Layer}
  \foreachitem \N \in \Nnod{ 
    \def\lay{\Ncnt} 
    \pgfmathsetmacro\prev{int(\Ncnt-1)} 
    \message{\lay,}
    \foreach \i [evaluate={\c=int(\i==\N); \y=\N/2-\i-\c*\yshift;
                 \index=(\i<\N?int(\i):"\Nstr[\lay]");
                 \x=\lay; \n=\nstyle;}] in {1,...,\N}{ 
      \node[node \n] (N\lay-\i) at (\x,\y) {$\Cstr[\lay]_{\index}$};
      
      \ifnum\lay>1 
        \foreach \j in {1,...,\Nnod[\prev]}{ 
          \draw[connect,white,line width=1.2] (N\prev-\j) -- (N\lay-\i);
          \draw[connect] (N\prev-\j) -- (N\lay-\i);
        }
      \fi 
    }
    \path (N\lay-\N) --++ (0,1+\yshift) node[midway,scale=1.5] {$\vdots$};
  }
  \node[above=5,align=center,black] at (N1-1.90) {Input\\[-0.2em]layer};
  \node[above=2,align=center,black] at (N3-1.90) {Hidden layers};
  \node[above=10,align=center,black] at (N\Nnodlen-1.90) {
  Output\\[-0.19em]layer};
\end{tikzpicture}
\caption{\centering 
Schematic representation of the NN used when fitting the $Q_\beta$. }
\label{fig:nn_architecture}
\end{figure}

Multilayer Perceptrons (MLPs) are a type of Artificial Neural Network (ANN) in 
which neurons are connected via a directed acyclic graph.
They have one or more inputs, more than one hidden layer, and one output 
layer. 
The hidden neurons simply propagate information forward to the next layer, 
each layer is a linear transformation of the previous one via 
non-linearities, which are required to allow a general approximation of a 
black box function such as the studied 
$\mathcal Q_\beta = \mathcal Q_\beta (A, Z, N,...)$. 
Such functions are referred to as the layer activation function and the 
output layer corresponds to a prediction of $Q_\beta$. 
The Neural Network (NN) architecture schema is shown in 
Fig.~\ref{fig:nn_architecture}. 

The Gaussian Error Linear Unit (GeLU) activation function was included as 
a non-linearity, given that it has been shown that helps reduce several 
problems associated with small datasets~\cite{nguyen2021analysis}. 
In addition, an inter-layer dropout is included, which discards random 
connections between layers with a probability of 0.1. 
This has been shown also to help in the optimization 
procedure~\cite{garbin2020dropout}.

Several experiments modifying the hyper-parameters via optimization 
methods were performed to obtain a suitable configuration of both the 
number of neurons per layer and the batch size~\cite{akiba2019optuna}. 
A proper learning rate was also obtained using the method of cyclical 
learning rates as proposed in Ref.~\cite{smith2017cyclical}. 
The training process used a decaying learning rate and was implemented 
using the PyTorch Lightning library \cite{Falcon_PyTorch_Lightning_2019}. 
This used mini-batch gradient descent to optimize the Mean Squared Error 
Loss (MSELoss) with the Adam optimizer~\cite{rajendra2021optimization}. 
Before starting the actual training process, three warm-up rounds were 
used.
%


\subsubsection{Gradient boosting machines}
Besides being increasingly known and used in different areas of physics, 
ANN are far from being the only available option for an explainable and 
accessible model, a clear example is an extreme gradient boosting (with 
\textsc{XGBoost}) algorithm~\cite{Chen_2016}, which uses the gradient 
boosted trees (BDT) method for fitting several tree regressors (or weak 
linear learners) using a regularized loss function. 
These are compared and later assembled with the approach of ML boosting. 
They perform especially well on tabular data and in most cases 
out-performing deep-learning approaches, requiring only fractions of the 
computational cost~\cite{dl_not_all}.
Here, the {XGBoost} model with the optimization of hyper-parameters 
with Optuna \cite{optuna} was implemented and several hyper-parameters
such as the depth, learning rate, and the number of learners were 
inspected. 
Moreover, callbacks to avoid over-fitting were added to the loss 
minimization process. 

\subsubsection{Attention Networks}
The \textit{attention mechanism}~\cite{vaswani2017attention} is implemented
here to properly learning weights and relations within the input via 
learning queries and keys. 
Besides, this approach was found to be successful for unstructured
data in several tasks such as protein unfolding and natural language 
processing, it had not been especially useful for tabular data until the 
the appearance of the TabNet model~\cite{tabnet}.
In this approach, the TabNet model was used with Weighted Adam optimizer 
and early stopping callbacks to avoid overfitting. 

\subsubsection{Ensemble model}
Even though the fitting generated with the methods above might fit 
accurately to the datasets, the models trained can generally either 
underfit, overfit or just be poorly configured. 
The aim of the ensemble method is to facilitate the best of several Base 
Models to train a strong Prediction Model. 
This is achieved via a Voting Ensemble Regression (VER), a linear 
combination of the predictions is done by weighting the individual models 
according to their performance on the validation split. 
Specifically, the weighting of the BDTs, the ANN, and the TabNet model is 
performed by evaluating both the RMSE and the standard deviation over a 
5-fold in the validation set.

\subsection{Data Augmentation Techniques\label{sec:usage_unc}}
From the initial evaluation of the feature importance, experimental 
uncertainties arose as noteworthy in the regression of $Q_\beta$. 
Here, they come from the combination of both epistemic and systematic 
errors. 
Also, a characteristic often makes it difficult for the models to improve 
their performance is the limited experimental data they are training on. 

Therefore, a method for obtaining more diverse and accurate data is useful. 
A Monte Carlo sampling on each one of the parameters with registered 
uncertainty is proposed. 
Assuming that each one of the variables has an error normally distributed 
and centered on the reported data. 
This allows obtaining a significant data set for further improving the ML 
models with increased robustness. 
This is only done in the training entries so that no isotopes are overlapped
between the split, ensuring no data leakage. 
Notice that this approach can be followed by the same procedure as an 
augmentation at testing time, as it has been shown to considerably improve 
robustness~\cite{test_time_aug}.

Following this procedure, the original training data was overpopulated, 
from $1912$ to a total of $573600$ samples, followed by the removal of 
the columns related to uncertainties, since this information has already 
been taken into account. 
The models that were trained on the original data (which will continue to 
be referred to as AME2020) were also trained and tuned on the augmented 
dataset.

\subsection{Periodic Feature Injection\label{sec:usage_priodic}}
In the spirit of the augmentation technique presented in 
Sect.~\ref{sec:usage_unc}, physics-inspired features were added to both 
datasets (i.e to the AME2020 and to the augmented dataset from 
Sect~\ref{sec:usage_unc}).
The periodic group was added to the data as a one-hot encoded feature. 
Moreover, the \textit{magic numbers} such that they are arranged into 
complete shells in the nucleus were also taken for each one of the 
isotopes, both for $N$ and $Z$. 
For this, the dataset includes the difference to the closes magic number 
(which we denote as $m_N$ and $m_Z$, respectively). 
In addition, the ratio $m_Z/(m_N+1)$ and $Z/N$ were also included.

Given that this method heavily relies on the current atomic model, we 
denote the generated data as \textit{Injected}.

\section{Results}
For each one of the models described above, the coefficient of 
determination ($R^2$) of the fit, the Mean Squared Error (MSE), the Root 
The Mean Squared Error (RMSE), and the Mean Absolute Error (MAE) were 
calculated.
We use this collection of metrics to get an estimation of how well 
distributed and big the deviations between $\mathcal{Q}_\beta$ and 
$Q_\beta$ is for the unseen testing dataset.
Below, are the results obtained using the approaches of different ML models 
with several datasets.

\subsubsection*{AME2020}
After fitting several regressors on the original split, modern architecture 
clearly outperforms the previous state-of-the-art RMSE. 
Notably, results obtained by the OLS are not considerably bigger compared 
to the baseline ANN. 
Moreover, experiments suggest that the number of parameters that have to be 
trained on the TabNet requires considerably more than one entry per 
isotope, thus the metrics are not taken into account. 

\begin{table}[h!]
\begin{center}
\small
\begin{tabular}{|lcccc|}
\hline \hline
\label{table:regression_metrics}
Model\  &$R^{2}$\ & MSE \begin{tiny}[MeV$^2$]\end{tiny}\ & RMSE 
\begin{tiny}[MeV]\end{tiny}\ & MAE \begin{tiny}[MeV]\end{tiny}  
\\
\hline \hline
\textbf{Baseline} & 0.926 & 4.134 & 2.033 & $1.518$
\\ 
\textbf{OLS} & 0.868 & $7.943$ & 2.818 & 2.043 
\\
\textbf{ANN} & 0.946 & $2.724$ & 1.650 & 1.289
\\
\textbf{BDT} & 0.958 & $2.271$ & 1.715 & 1.275
\\
\textbf{Ensemble} & 0.957 & $2.305$ & 1.518 & 1.191
\\
\hline
\hline
\end{tabular}
\\
\caption{\centering Regression metrics for the different classifiers 
trained on the original AME2020 training set.}
\end{center}
\end{table}

\begin{figure}[!h]
    \centering\label{fig:results_ensemble_naive_qb}
    \includegraphics[width=0.5\textwidth]
    {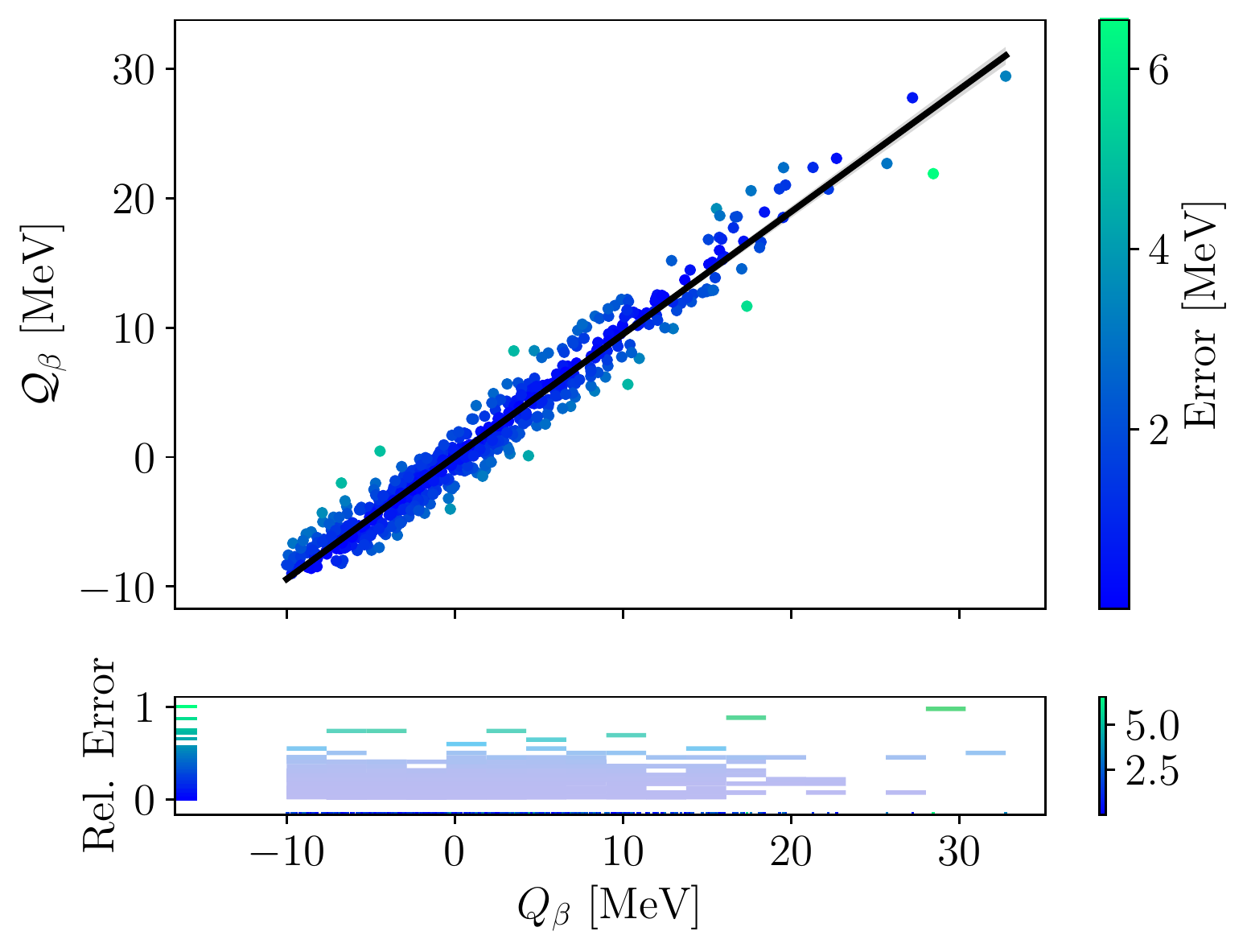}
    \caption{\centering Residuals of the ensemble of models trained on 
    the original AME2020 dataset.}
\end{figure}

\begin{figure}[!h]
    \centering\label{fig:ensemble_naive_error_NZ}
    \includegraphics[width=0.5\textwidth]
    {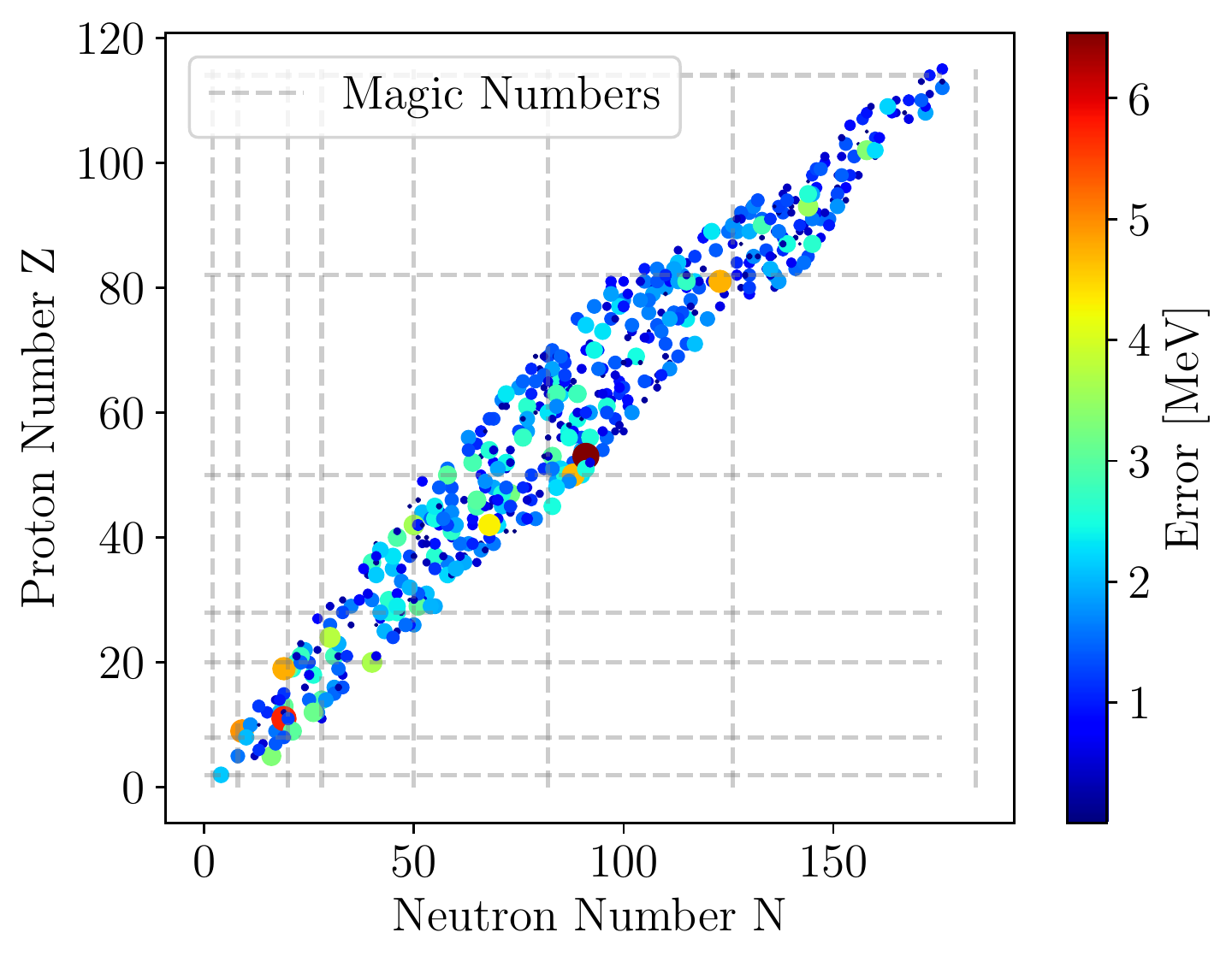}
    \caption{\centering Distribution of the errors in the predicted 
    $\mathcal{Q}_\beta$ by the models trained in the original AME2020 
    dataset within the $N$ and $Z$ plane, the size of the points 
    represented is also determined by the deviation in $Q_\beta$.}
\end{figure}

Besides the distribution of errors seems to be uniformly distributed for
different $Q_\beta$ within the dataset 
(cf~\ref{fig:results_ensemble_naive_qb}), it is remarkable that the 
distribution of outliers is not with a clear higher deviation on low 
number of nucleons, as illustrated in Fig~\ref{fig:ensemble_naive_error_NZ} 
and more notably, on the intersection points for the atomic magic numbers.

Furthermore, the coefficients extracted by the OLS regression reflect the 
fact of not only interference but also of irrelevance for some fitted 
parameters, evidenced in Table~\ref{table:regression_res}.

\begin{table}[h!]
\begin{tabular}{|lcccc|}
\hline \hline
\small
\label{table:regression_res}
Variable &  $c_i$ \ \ &  $\sigma$ &  $P>\|t\|$ & C.I. @ $95\%$ 
\\
\hline \hline
N            &  533.02 &   10.91  &  0.00 &   [511.60 ,  554.43] 
\\
Z            & -925.62 &   16.08  &  0.00 &  [-957.19 ,  -894.06] 
\\
ME        &   -0.00 &    0.01 &  0.97 &    [-0.01 ,  0.01] 
\\
$\delta_{ME}$    & -196.84 &  578.31  &  0.73 & [-1331.34 ,   937.66] 
\\
$E_{b}$     &    1.21 &    0.04 &  0.00 &     [1.13 ,     1.30] 
\\
$\delta_{E_b}$  &  263.86 &   20.87  &  0.00 &   [222.91 ,   304.80] 
\\
AM &   -0.001 &    0.00 &  -2.01 &    [-0.01 ,    -0.00] 
\\
$\delta_{AM}$      &  186.21 &  538.77 &   0.73 &  [-870.73 ,  1243.14] 
\\

\hline \hline
\end{tabular}
\caption{Results on the fitted coefficients for the OLS model. 
$c_i$ is the coefficient and corresponds to the rate of change of 
$Q_\beta$ with respect to each variable, $\sigma$ is the standard error,
$P>\|t\|$ is the probability that the coefficient is measured, and 
C.I. @ $95\%$ is representing the range of the coefficients within 
95$\%$ confidence level.} 
\end{table}

\subsubsection*{Augmented Dataset}
Since the amount of data in this used in this approach is considerably big,
the OLS was not implemented. 

\begin{table}[h!]
\begin{center}
\small
\begin{tabular}{|lcccr|}
\hline \hline
\label{table:regression_metrics}
Model\  &$R^{2}$\ & MSE \begin{tiny}[MeV$^2$]\end{tiny}\ & RMSE 
\begin{tiny}[MeV]\end{tiny}\ & MAE \begin{tiny}[MeV]\end{tiny}  \\
\hline \hline
\textbf{Baseline} & 0.926 & 4.134 & 2.033 & $1.518$
\\ 
\textbf{ANN} & 0.964 & $1.895$ & 1.372 & 1.084
\\
\textbf{TabNet} & 0.958 &$ 0.995$ & $1.149$ & 0.987
\\
\textbf{BDT} & 0.986 &$ 0.749$ & $0.865$ & 0.371
\\
\textbf{Ensemble} & 0.987 &$ 0.686$ & $0.828$ & 0.460
\\
\hline \hline
\end{tabular}
\\
\caption{\centering 
Regression metrics for both the original AME2020 and the augmented data.}
\end{center}
\end{table}

\begin{figure}[!h]
    \centering
    \includegraphics[width=0.5\textwidth]
    {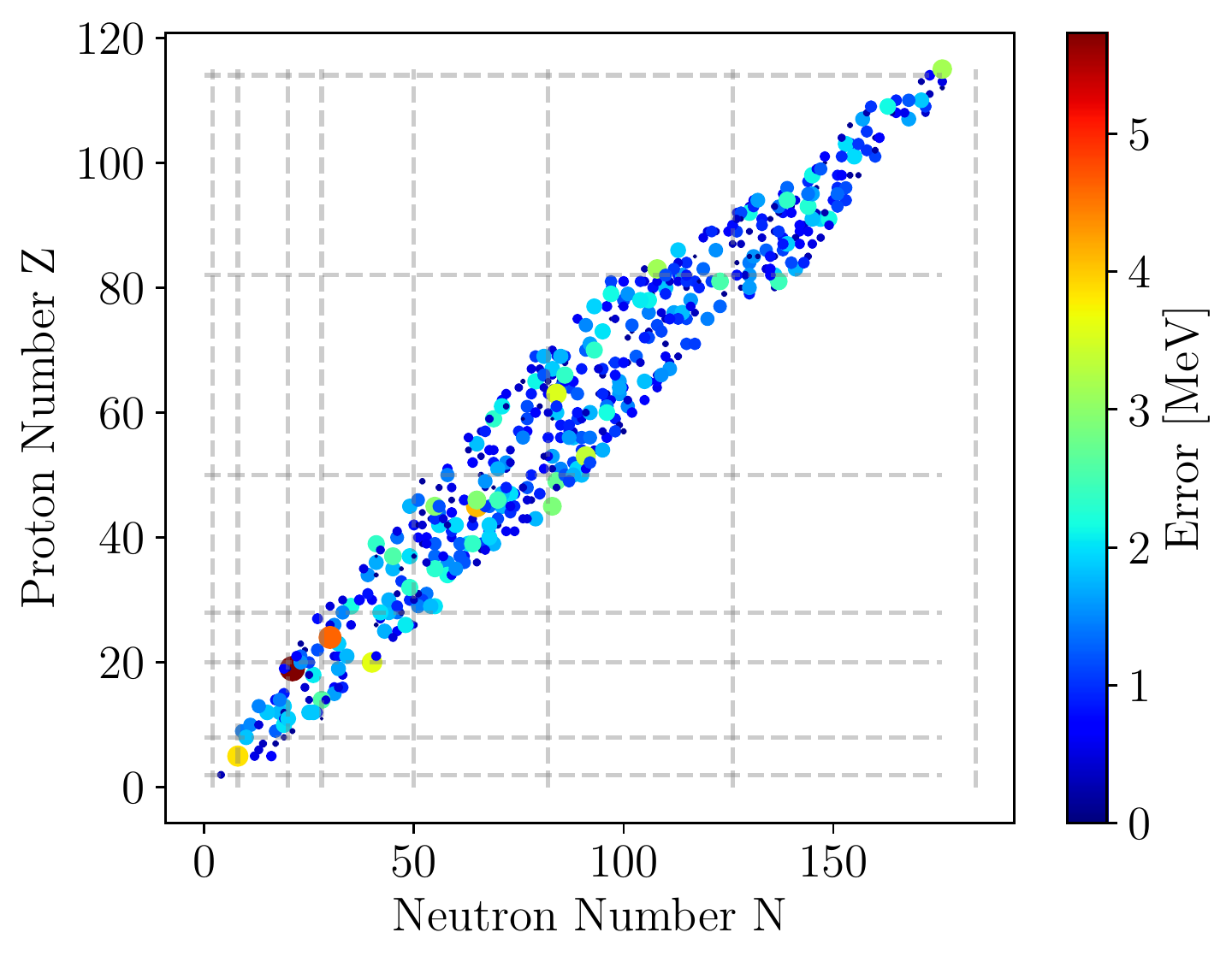}
    \caption{\centering Real $Q_\beta$ versus predicted 
    $\mathcal{Q}_\beta$ for the ensemble model
    trained on the augmented dataset.}
    \label{fig:results_ensemble}
\end{figure}

All the models were trained on the same test split that was established 
from the beginning of the process. And the good performance with no 
manifestations of overfitting shows that the models are capable of 
extrapolating to unseen atomic feature configurations.
Remarkably, the ensemble model achieves remarkable accuracy with 
fewer error outliers than all of the other models, taking advantage of 
the uncertainties with the original AME2020 dataset and the augmentation 
process. 
Moreover, the voting based on splits approach allows to effectively target 
robustness, as the ensemble method achieved a $12\%$ lower maximum 
deviation in $Q_\beta$ over the test isotopes.
This allows for state-of-the-art metrics and a remarkably low 
MSE, as shown in Fig.~\ref{fig:results_ensemble} for the case 
of the ensemble model.

{}

\begin{figure}[!h]
    \centering
    \includegraphics[width=0.5\textwidth]
    {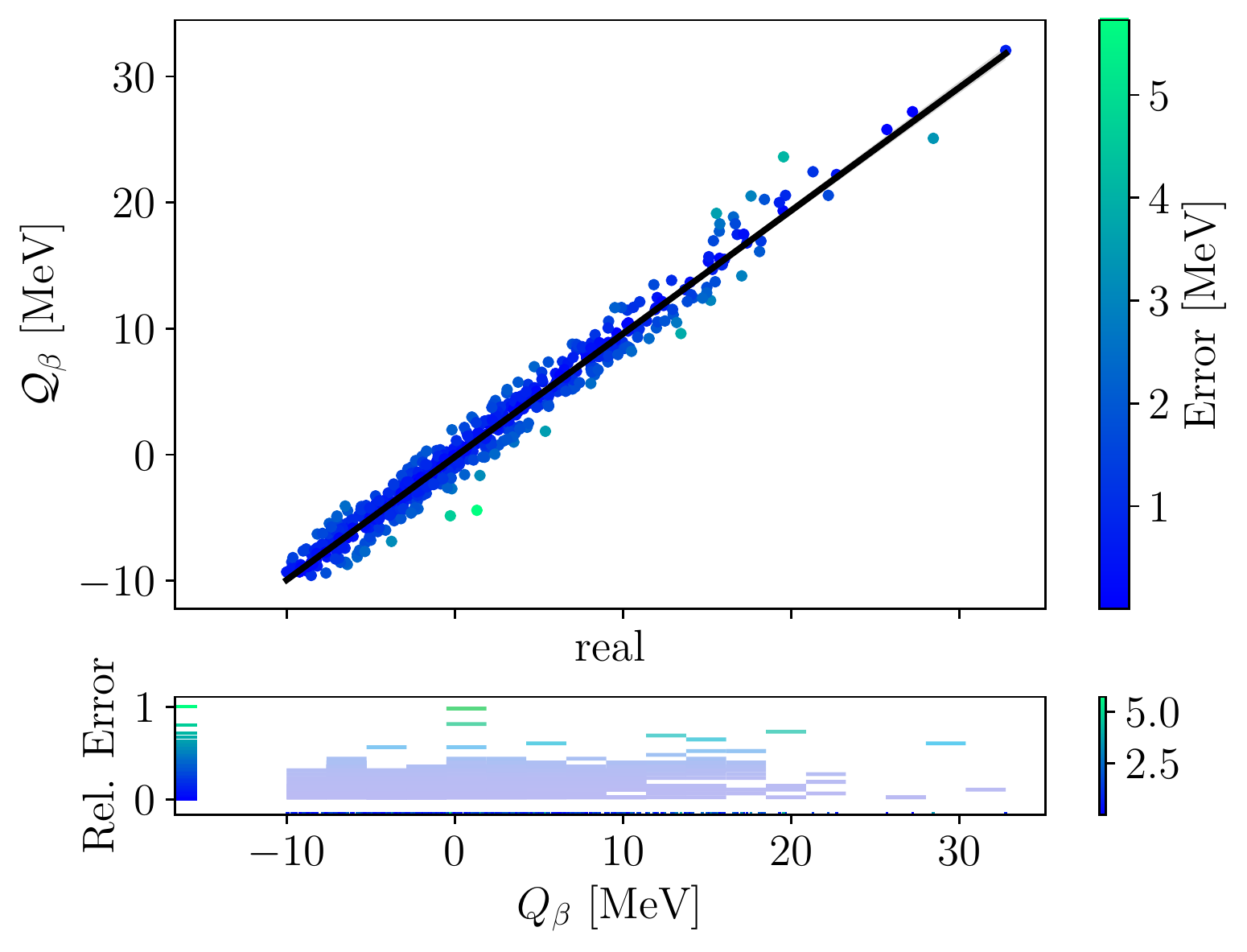}
    \caption{Deviation in Predicted $Q_\beta$ for the test isotopes by 
    the models trained on the augmented dataset.}
    \label{fig:ensemble_augmented_qb}
\end{figure}

Furthermore, besides the error corresponding to isotopes near magic 
numbers still having higher errors, the deviations were reduced uniformly 
for test isotopes, as illustrated in Fig.~\ref{fig:ensemble_augmented_qb}. 

\subsubsection*{Injected Dataset}
Within the same nature, the models trained on the dataset containing the 
periodic group, and information related to the magic numbers, as presented 
in Section ~\ref{sec:usage_priodic}. 
The first experiments showed that the encoded periodic group is detrimental 
to the performance, and thus those features were dropped.
This approach was implemented both on the Augmented and in the original 
AME2020 datasets, both of which are presented in 
Table~\ref{table:regression_metrics_injected}.

\begin{table}[h!]
\begin{center}
\small
\begin{tabular}{|lcccr|}
\hline \hline
\label{table:regression_metrics_injected}
Model\  &$R^{2}$\ & MSE \begin{tiny}[MeV$^2$]\end{tiny}\ & RMSE 
\begin{tiny}[MeV]\end{tiny}\ & MAE \begin{tiny}[MeV]\end{tiny}  
\\
\hline \
AME2020 & &  &   & 
\\
\hline 
\textbf{Baseline} & 0.926 & 4.134 & 2.033 & $1.518$
\\ 
\textbf{ANN} & 0.950 & $2.583$ & 1.607 & 1.252
\\
\textbf{BDT} & 0.981 & $0.756$ & 0.869 & 0.613
\\
\hline
Augmented & & & & 
\\
\hline
\textbf{ANN} & 0.970 & $1.640$ & 1.281 & 1.031
\\
\textbf{TabNet} & 0.982 &$ 0.993$ & $0.997$ & 0.774
\\
\textbf{BDT} & 0.989 &$ 0.576$ & $0.759$ & 0.534
\\
\textbf{Ensemble} & 0.991 &$ 0.511$ & $0.714$ & 0.510
\\
\hline \hline
\end{tabular}
\\
\caption{\centering 
Regression metrics for both the original AME2020 and the augmented data
after performing feature Injection.}
\end{center}
\end{table}

This approach demonstrates the importance of generating features based on
theoretical models, as higher-level atomic features clearly allow models 
to fit the function better.
This is noticeable in the distribution of deviations.

\begin{figure}
    \centering
    \includegraphics[width=0.5\textwidth]
    {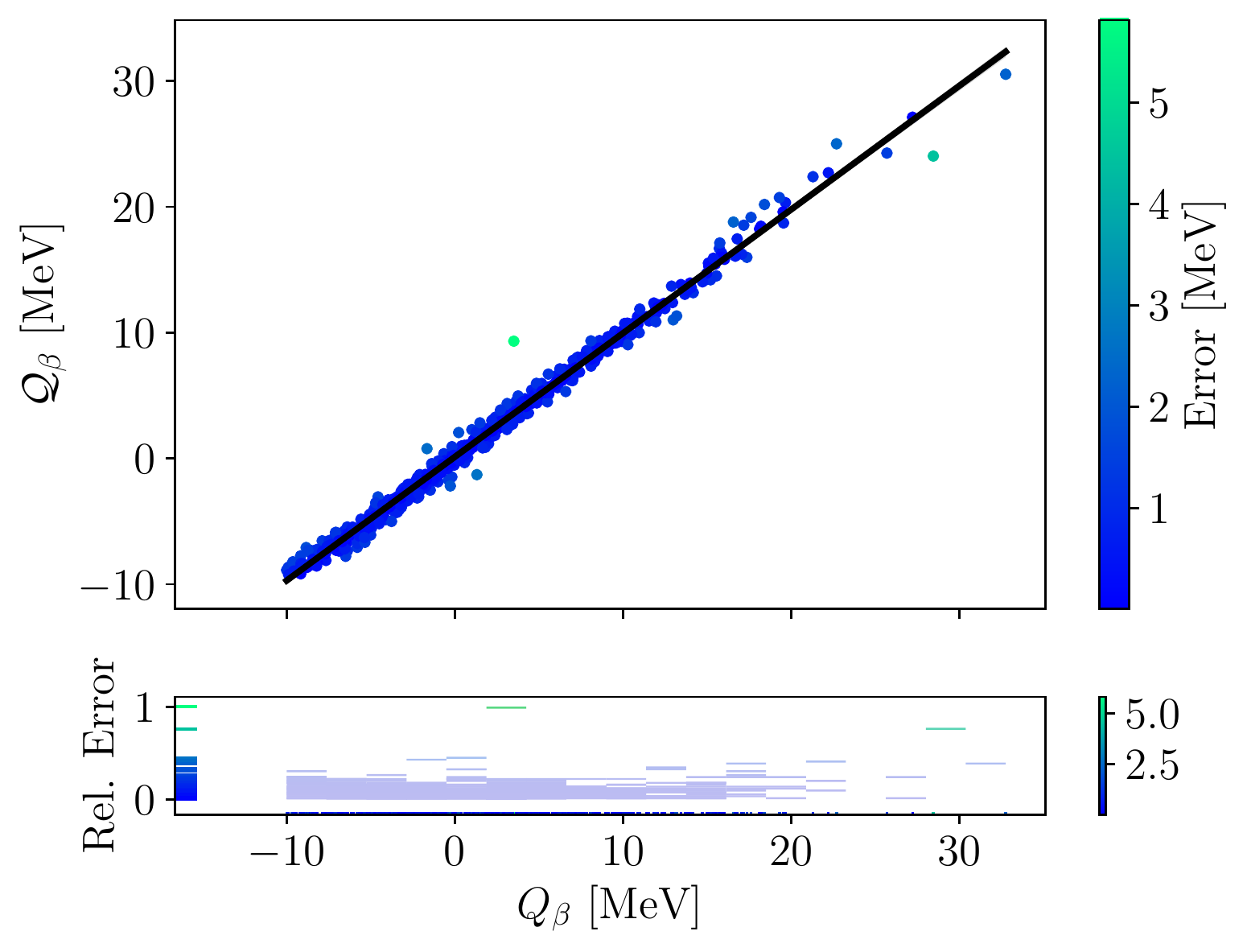}
    \caption{\centering Real $Q_\beta$ versus predicted 
    $\mathcal{Q}_\beta$ by the ensemble model using the injected 
    periodic features.}
    \label{fig:my_label}
\end{figure}

\begin{figure}
    \centering
    \includegraphics[width=0.5\textwidth]
    {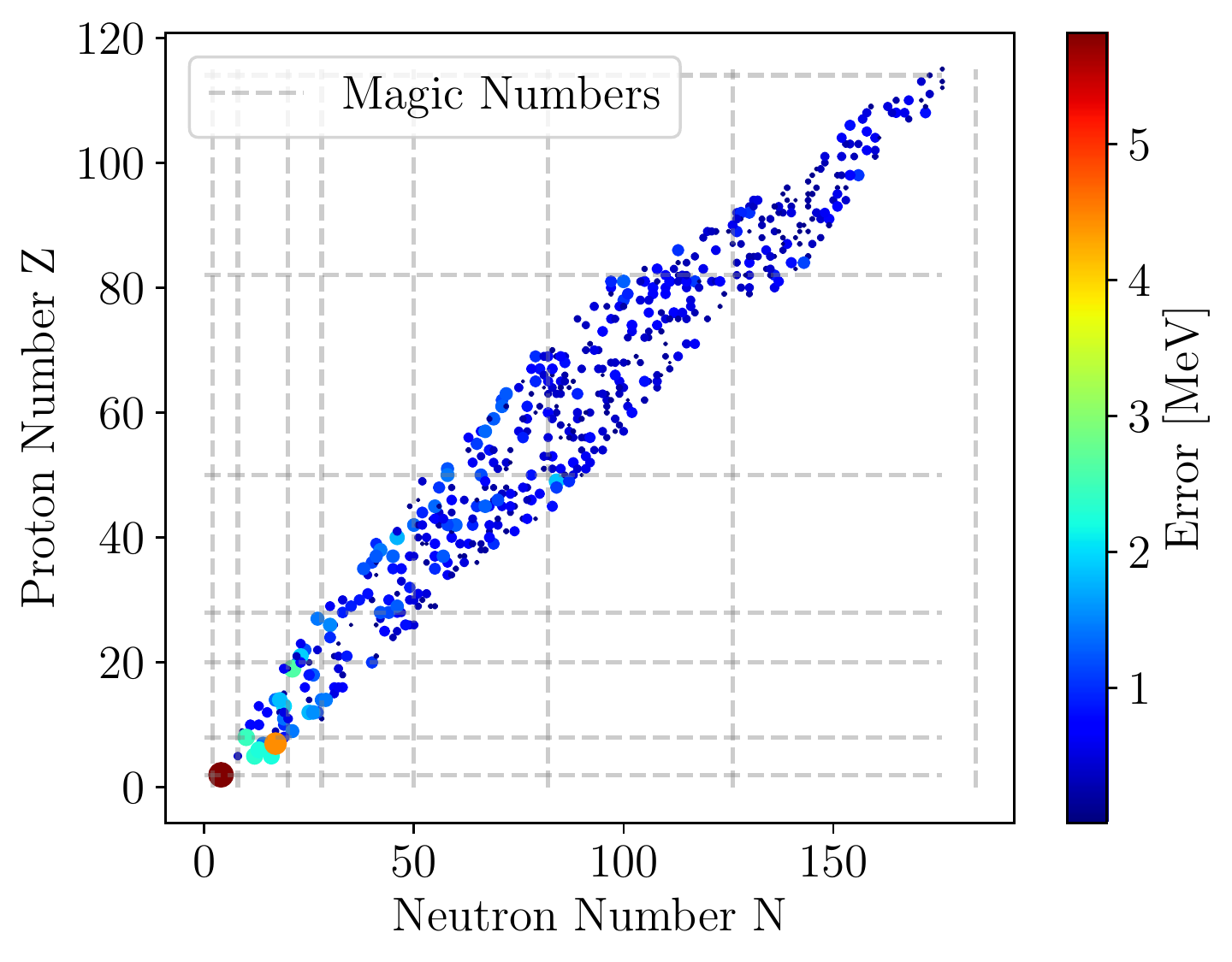}
    \caption{\centering Distribution of deviations in $Q_\beta$ as 
    predicted by the ensemble of all models on the testing isotopes with
    the periodic injected features.}
    \label{fig:ensemble_inj_error_NZ}
\end{figure}

As expected, this approach allows a considerable reduction in errors,
especially for the critical isotopes in not injected approaches, and with 
the highest error at a low number of isotopes, as illustrated in 
Fig.~\ref{fig:ensemble_inj_error_NZ}. 
Notably, the two isotopes with an error above the percentile 90 correspond 
to Helium (isotope with $A=6$) and Nitrogen (isotope with $A=24$). 

\section{Explaining the Black-Box}
\begin{figure}[!h]
    \centering
    \includegraphics[width=0.5\textwidth]
    {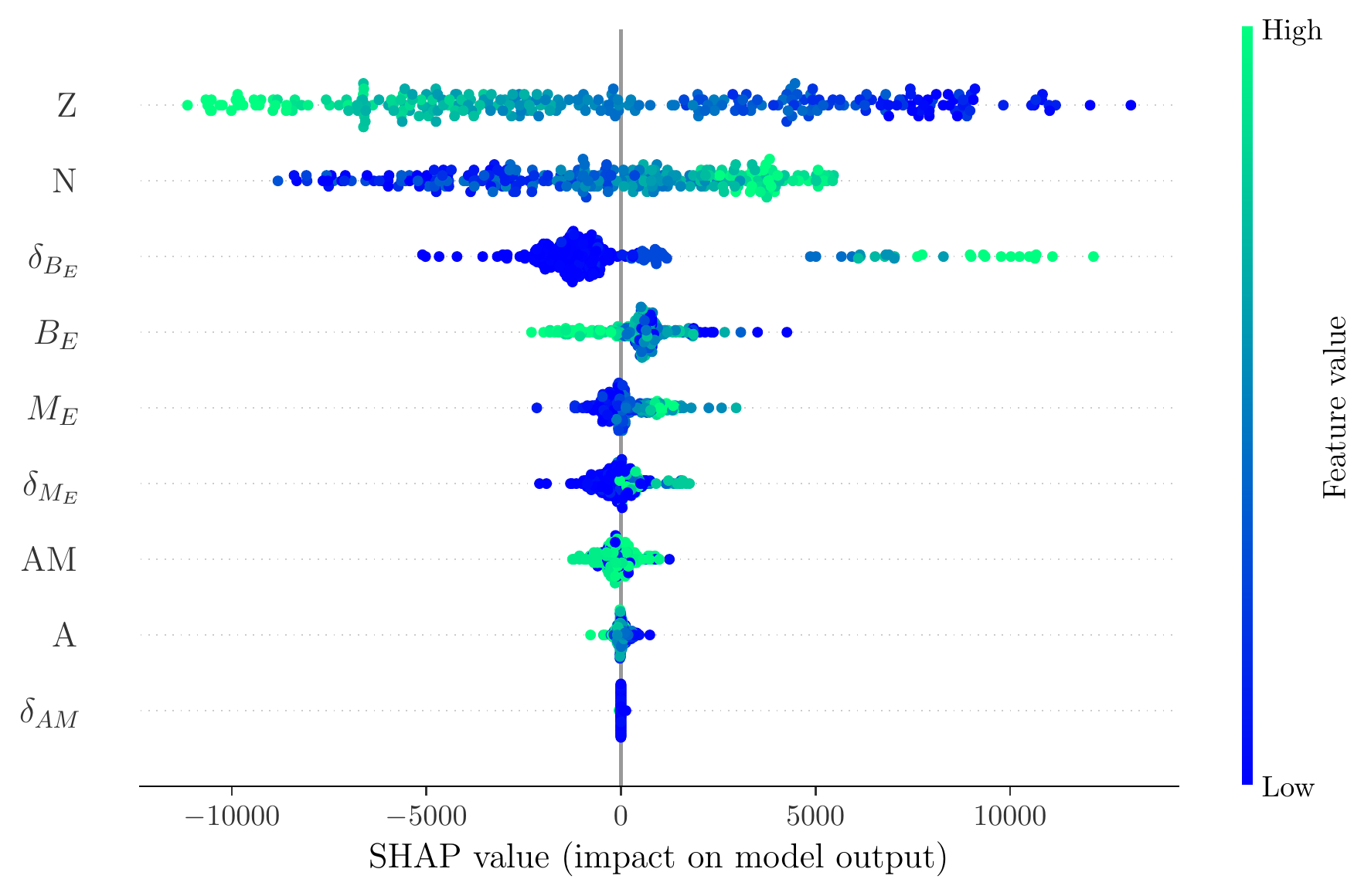}
    \caption{\centering Feature importance of the hyper-tuned BDT on the 
    AME2020 dataset.}
    \label{fig:shaps}
\end{figure}

Up to this point, ML techniques have been shown to provide a highly 
accurate regression of the $\beta$-decay energies.
However, it is possible further to inspect the impact of each one of the 
features on the output of the model. 
This can be done via the evaluation of the Shapley Additive explanations 
(SHAP), a model-agnostic explainability method inherited from game theory, 
where the feature importance of the feature is calculated via a global 
analysis of the marginals per each input feature~\cite{NIPS2017_7062}. 
For this end, the KernelSHAP technique implemented in 
Ref.~\cite{lundberg2020local2global} is applied. 
The results are presented in Fig.~\ref{fig:shaps} for the case of BDT, 
which was fully trained and hyper-optimized on the AME2020 training data.

In accordance with our explorations with the PCA approach, the 
uncertainties play a significant role in explaining the model output and 
that in general, the mass excess is considered charged. 
The feature importance after training on the augmented 
the dataset was also studied.
It was found that the importance scores for the physical features ($N$, 
$Z$, Mass Excess, $B_E$, Atomic Mass, and $A$) kept their assigned 
importance as expected from an orthogonal permutational scoring. 
Moreover, the feature weights inside the trees for the 
BDT and the feature permutation importance for the ANN, which matches the 
overall order was also investigated.

The models thus fit the coefficients with a major importance of $Z$ and 
$N$, which could in fact explain the good results obtained in 
Ref.~\cite{DBLP:journals/corr/abs-1907-10902} with such a simple 
architecture. 
Furthermore, the relation found by the black boxes, in general, is 
inversely proportional for $Z$ and direct for $N$.

\begin{figure}[!h]
    \centering
    \includegraphics[width=0.5\textwidth]{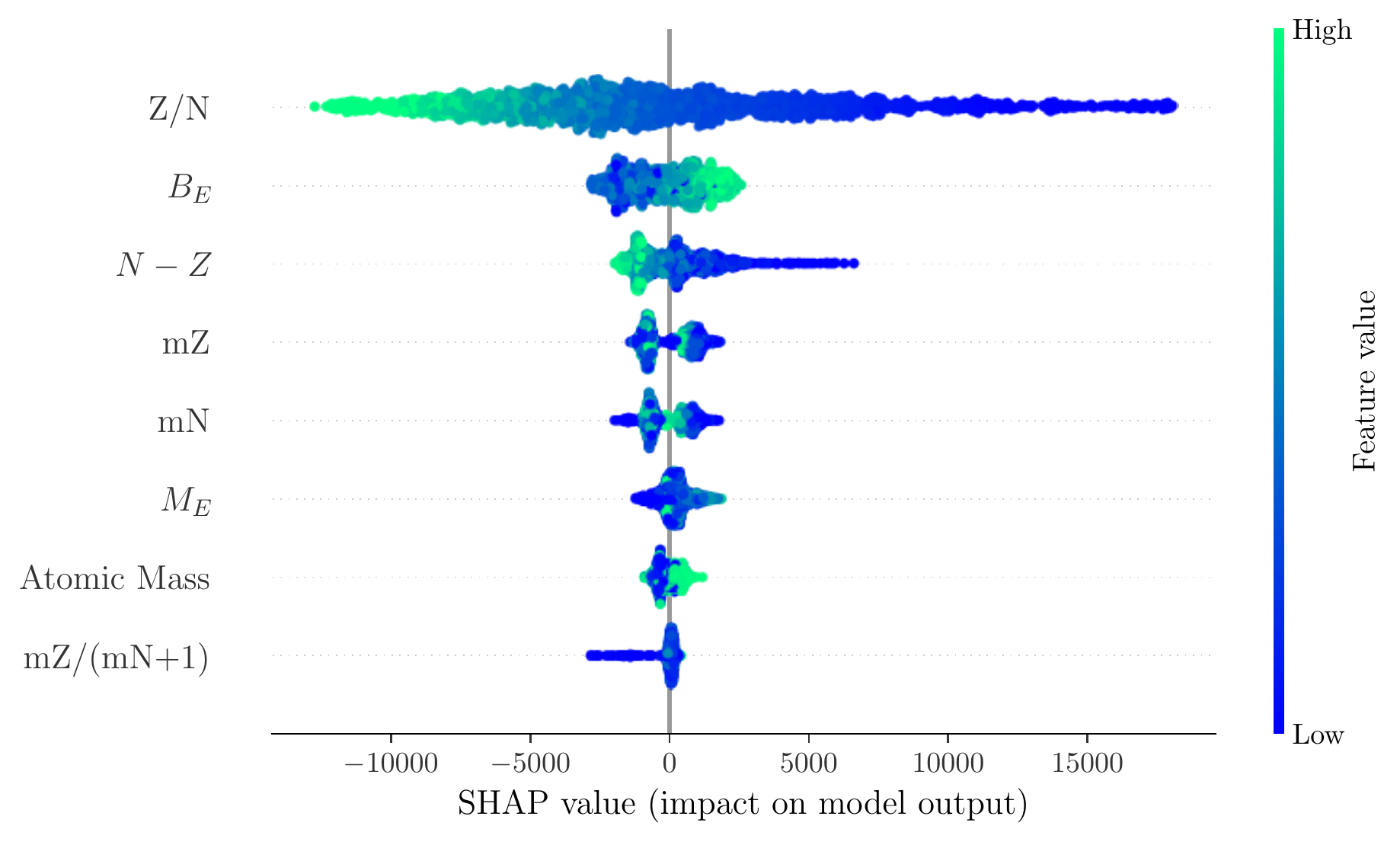}
    \caption{\centering Permutational feature importance for the model 
    trained on the augmented set with injected features, only the 8 most
    important features are displayed.}
    \label{fig:f_scores}
\end{figure}

More importantly, the fact that the groups were detrimental to the 
performance of the regressors was validated by the permutational scoring.
By considering the periodic injected features, the high importance of 
the ratios $N/Z$, and the magic numbers give insights into why the 
critical isotopes near the magic numbers were predicted with lower errors 
(cf ~\ref{fig:f_scores}). 

\begin{figure}[!h]
    \centering
    \includegraphics[scale=0.55]{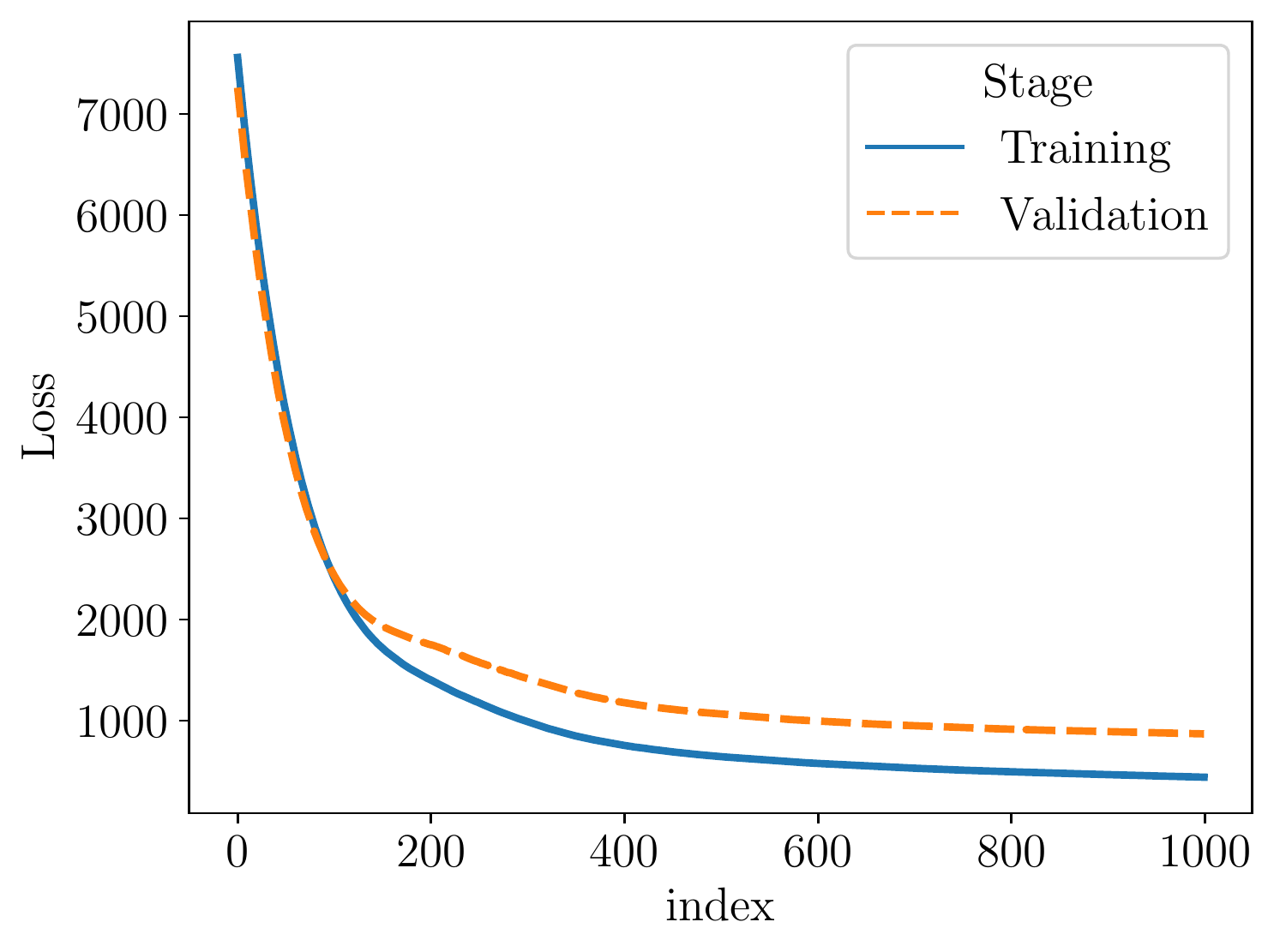}
    \caption{\centering MSE Loss of the BDT model after hyper-tuning 
    trained on the augmented dataset with feature injection, .}
    \label{fig:losses_xgb}
\end{figure}

Remarkably, the feature importance averaged over 5 runs for the models 
show to have a relative
low $F$ score, experiments show that injecting periodic variables allows 
the BDT model to achieve the same metrics as the augmented dataset, and to 
increase metrics on the augmented dataset with virtually no overfitting, as 
illustrated in Fig.~\ref{fig:losses_xgb}.

\begin{figure}[!h]
    \centering
    \includegraphics[width=0.5\textwidth]{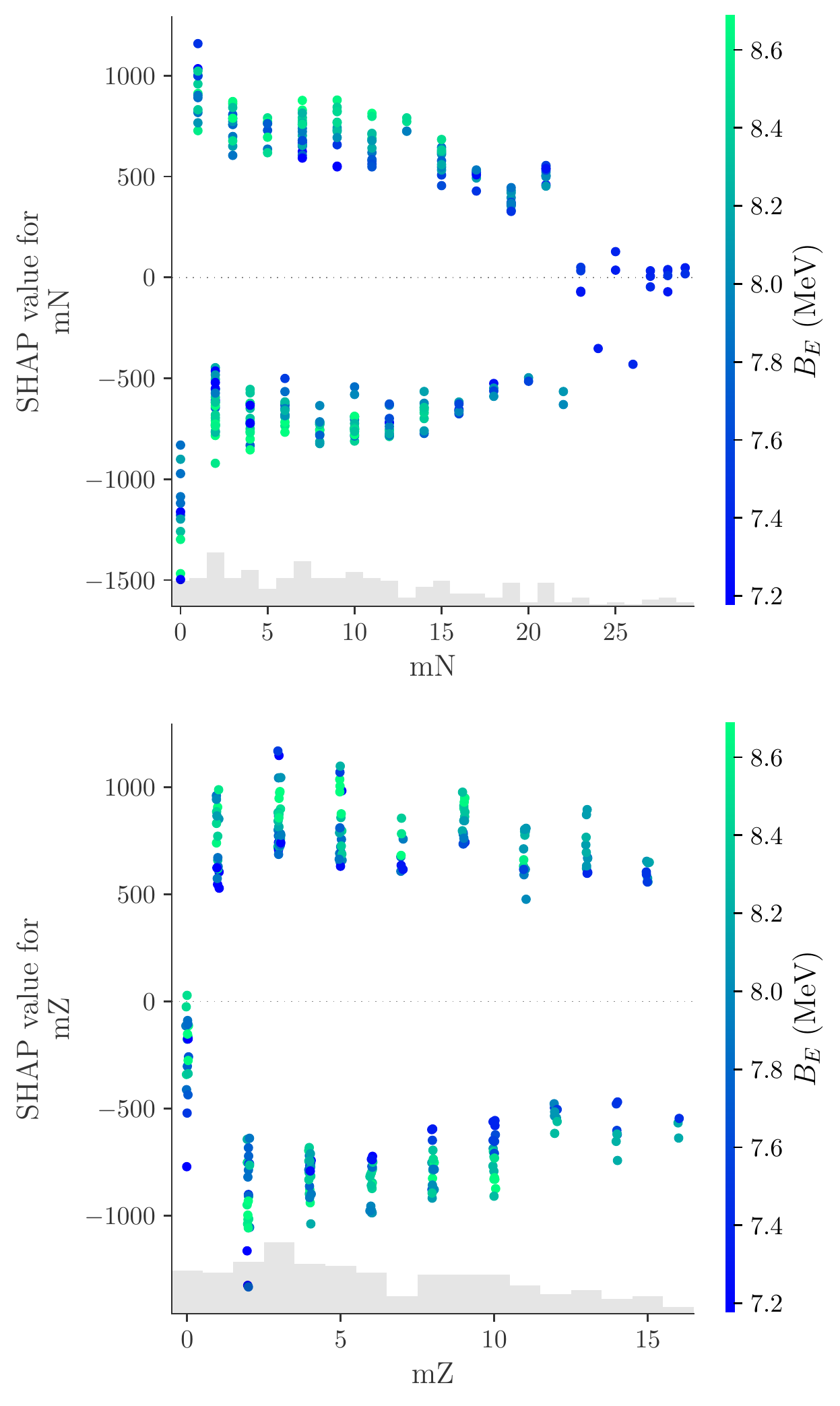}
    \caption{\centering Distribution of the feature importance of $mZ$ 
    and $mN$ where     the alternating pattern in parity is evident.}
    \label{fig:altern_shap}
\end{figure}

Further inspection of the injected features also shows interesting behavior 
arising from physical phenomena. 
For instance, both for $mZ$ (absolute difference between $Z$ and the 
closest magic number) and $mN$ (difference between $N$ and the closest
magic number) have alternating feature importance with parity as shown in 
Fig~\ref{fig:altern_shap}.
The oddness in the number of nucleons in relation to the magic numbers,
in general, results in lower $B_E$, which is in excellent agreement with
the Nuclear Shell Model \cite{brown1988status} as well as theoretical 
approximations of nuclear structure such as the Weizsäcker 
formula~\cite{lilley2013nuclear}.

This alternation in the SHAP importance value is also larger near the magic 
numbers, which might explain the improvement in deviations near magic 
numbers with the feature injection, and is clear evidence that the model 
is already capable of getting insights into the nuclear structures.

\section{Conclusions and Outlooks}
In this work, an ML model was developed to predict the $\beta$-decay
energies of isotopes using the experimental dataset AME2020 
\cite{AME2020}. 
Using modern techniques and data augmentation strategies, it was shown 
that our models can considerably decrease the error in estimating 
$Q_\beta$ compared with previous ML approaches while maintaining 
interpretability.
The importance of considering experimental uncertainties was illustrated 
as the models incorporating sampling techniques become considerably more 
robust to noise. 
Furthermore, it was demonstrated that by incorporating physical features
such as magic numbers, models can be highly improved. Using explainability
tools it is shown that the models have learned fundamental features of the
atomic structure.

We look forward to generating clearer physical intuition from the 
explainability extracted from several ML techniques. 
Clearly, the intersection between ML and physics is only starting to 
showcase its potential.
%
\section{Acknowledgments}

This work was supported by the Erasmus+ program and the R+D+I 
efforts from guane enterprises.

\newpage

\bibliography{references}

\begin{thebibliography}{51}%
\makeatletter
\providecommand \@ifxundefined [1]{%
 \@ifx{#1\undefined}
}%
\providecommand \@ifnum [1]{%
 \ifnum #1\expandafter \@firstoftwo
 \else \expandafter \@secondoftwo
 \fi
}%
\providecommand \@ifx [1]{%
 \ifx #1\expandafter \@firstoftwo
 \else \expandafter \@secondoftwo
 \fi
}%
\providecommand \natexlab [1]{#1}%
\providecommand \enquote  [1]{``#1''}%
\providecommand \bibnamefont  [1]{#1}%
\providecommand \bibfnamefont [1]{#1}%
\providecommand \citenamefont [1]{#1}%
\providecommand \href@noop [0]{\@secondoftwo}%
\providecommand \href [0]{\begingroup \@sanitize@url \@href}%
\providecommand \@href[1]{\@@startlink{#1}\@@href}%
\providecommand \@@href[1]{\endgroup#1\@@endlink}%
\providecommand \@sanitize@url [0]{\catcode `\\12\catcode `\$12\catcode
  `\&12\catcode `\#12\catcode `\^12\catcode `\_12\catcode `\%12\relax}%
\providecommand \@@startlink[1]{}%
\providecommand \@@endlink[0]{}%
\providecommand \url  [0]{\begingroup\@sanitize@url \@url }%
\providecommand \@url [1]{\endgroup\@href {#1}{\urlprefix }}%
\providecommand \urlprefix  [0]{URL }%
\providecommand \Eprint [0]{\href }%
\providecommand \doibase [0]{http://dx.doi.org/}%
\providecommand \selectlanguage [0]{\@gobble}%
\providecommand \bibinfo  [0]{\@secondoftwo}%
\providecommand \bibfield  [0]{\@secondoftwo}%
\providecommand \translation [1]{[#1]}%
\providecommand \BibitemOpen [0]{}%
\providecommand \bibitemStop [0]{}%
\providecommand \bibitemNoStop [0]{.\EOS\space}%
\providecommand \EOS [0]{\spacefactor3000\relax}%
\providecommand \BibitemShut  [1]{\csname bibitem#1\endcsname}%
\let\auto@bib@innerbib\@empty
\bibitem [{\citenamefont {Alkhazov}\ \emph {et~al.}(1993)\citenamefont
  {Alkhazov}, \citenamefont {Batist}, \citenamefont {Bykov}, \citenamefont
  {Moroz}, \citenamefont {Orlov}, \citenamefont {Tarasov},\ and\ \citenamefont
  {Wittmann}}]{alkhazov1993beta}%
  \BibitemOpen
  \bibfield  {author} {\bibinfo {author} {\bibfnamefont {G.}~\bibnamefont
  {Alkhazov}}, \bibinfo {author} {\bibfnamefont {L.}~\bibnamefont {Batist}},
  \bibinfo {author} {\bibfnamefont {A.}~\bibnamefont {Bykov}}, \bibinfo
  {author} {\bibfnamefont {F.}~\bibnamefont {Moroz}}, \bibinfo {author}
  {\bibfnamefont {S.~Y.}\ \bibnamefont {Orlov}}, \bibinfo {author}
  {\bibfnamefont {V.}~\bibnamefont {Tarasov}}, \ and\ \bibinfo {author}
  {\bibfnamefont {V.}~\bibnamefont {Wittmann}},\ }\href@noop {} {\bibfield
  {journal} {\bibinfo  {journal} {Zeitschrift f{\"u}r Physik A Hadrons and
  Nuclei}\ }\textbf {\bibinfo {volume} {344}},\ \bibinfo {pages} {425}
  (\bibinfo {year} {1993})}\BibitemShut {NoStop}%
\bibitem [{\citenamefont {Kolesnikov}\ and\ \citenamefont
  {Krylova}(1959)}]{kolesnikov1959proton}%
  \BibitemOpen
  \bibfield  {author} {\bibinfo {author} {\bibfnamefont {N.}~\bibnamefont
  {Kolesnikov}}\ and\ \bibinfo {author} {\bibfnamefont {A.}~\bibnamefont
  {Krylova}},\ }\href@noop {} {\bibfield  {journal} {\bibinfo  {journal} {Zhur.
  Eksptl'. i Teoret. Fiz.}\ }\textbf {\bibinfo {volume} {37}} (\bibinfo {year}
  {1959})}\BibitemShut {NoStop}%
\bibitem [{\citenamefont {Carleo}\ \emph {et~al.}(2019)\citenamefont {Carleo},
  \citenamefont {Cirac}, \citenamefont {Cranmer}, \citenamefont {Daudet},
  \citenamefont {Schuld}, \citenamefont {Tishby}, \citenamefont
  {Vogt-Maranto},\ and\ \citenamefont {Zdeborov\'a}}]{Carleo:2019ptp}%
  \BibitemOpen
  \bibfield  {author} {\bibinfo {author} {\bibfnamefont {G.}~\bibnamefont
  {Carleo}}, \bibinfo {author} {\bibfnamefont {I.}~\bibnamefont {Cirac}},
  \bibinfo {author} {\bibfnamefont {K.}~\bibnamefont {Cranmer}}, \bibinfo
  {author} {\bibfnamefont {L.}~\bibnamefont {Daudet}}, \bibinfo {author}
  {\bibfnamefont {M.}~\bibnamefont {Schuld}}, \bibinfo {author} {\bibfnamefont
  {N.}~\bibnamefont {Tishby}}, \bibinfo {author} {\bibfnamefont
  {L.}~\bibnamefont {Vogt-Maranto}}, \ and\ \bibinfo {author} {\bibfnamefont
  {L.}~\bibnamefont {Zdeborov\'a}},\ }\href {\doibase
  10.1103/RevModPhys.91.045002} {\bibfield  {journal} {\bibinfo  {journal}
  {Rev. Mod. Phys.}\ }\textbf {\bibinfo {volume} {91}},\ \bibinfo {pages}
  {045002} (\bibinfo {year} {2019})},\ \Eprint
  {http://arxiv.org/abs/1903.10563} {arXiv:1903.10563 [physics.comp-ph]}
  \BibitemShut {NoStop}%
\bibitem [{\citenamefont {Karniadakis}\ \emph {et~al.}(2021)\citenamefont
  {Karniadakis}, \citenamefont {Kevrekidis}, \citenamefont {Lu}, \citenamefont
  {Perdikaris}, \citenamefont {Wang},\ and\ \citenamefont
  {Yang}}]{karniadakis2021physics}%
  \BibitemOpen
  \bibfield  {author} {\bibinfo {author} {\bibfnamefont {G.~E.}\ \bibnamefont
  {Karniadakis}}, \bibinfo {author} {\bibfnamefont {I.~G.}\ \bibnamefont
  {Kevrekidis}}, \bibinfo {author} {\bibfnamefont {L.}~\bibnamefont {Lu}},
  \bibinfo {author} {\bibfnamefont {P.}~\bibnamefont {Perdikaris}}, \bibinfo
  {author} {\bibfnamefont {S.}~\bibnamefont {Wang}}, \ and\ \bibinfo {author}
  {\bibfnamefont {L.}~\bibnamefont {Yang}},\ }\href@noop {} {\bibfield
  {journal} {\bibinfo  {journal} {Nature Reviews Physics}\ }\textbf {\bibinfo
  {volume} {3}},\ \bibinfo {pages} {422} (\bibinfo {year} {2021})}\BibitemShut
  {NoStop}%
\bibitem [{\citenamefont {Hastie}\ \emph {et~al.}(2009)\citenamefont {Hastie},
  \citenamefont {Tibshirani}, \citenamefont {Friedman},\ and\ \citenamefont
  {Friedman}}]{hastie2009elements}%
  \BibitemOpen
  \bibfield  {author} {\bibinfo {author} {\bibfnamefont {T.}~\bibnamefont
  {Hastie}}, \bibinfo {author} {\bibfnamefont {R.}~\bibnamefont {Tibshirani}},
  \bibinfo {author} {\bibfnamefont {J.~H.}\ \bibnamefont {Friedman}}, \ and\
  \bibinfo {author} {\bibfnamefont {J.~H.}\ \bibnamefont {Friedman}},\
  }\href@noop {} {\emph {\bibinfo {title} {The elements of statistical
  learning: data mining, inference, and prediction}}},\ Vol.~\bibinfo {volume}
  {2}\ (\bibinfo  {publisher} {Springer},\ \bibinfo {year} {2009})\BibitemShut
  {NoStop}%
\bibitem [{\citenamefont {Chen}\ \emph
  {et~al.}(2018{\natexlab{a}})\citenamefont {Chen}, \citenamefont {Liu},\ and\
  \citenamefont {Feng}}]{chen2018universal}%
  \BibitemOpen
  \bibfield  {author} {\bibinfo {author} {\bibfnamefont {C.~P.}\ \bibnamefont
  {Chen}}, \bibinfo {author} {\bibfnamefont {Z.}~\bibnamefont {Liu}}, \ and\
  \bibinfo {author} {\bibfnamefont {S.}~\bibnamefont {Feng}},\ }\href@noop {}
  {\bibfield  {journal} {\bibinfo  {journal} {IEEE transactions on neural
  networks and learning systems}\ }\textbf {\bibinfo {volume} {30}},\ \bibinfo
  {pages} {1191} (\bibinfo {year} {2018}{\natexlab{a}})}\BibitemShut {NoStop}%
\bibitem [{\citenamefont {Scarselli}\ and\ \citenamefont
  {Tsoi}(1998)}]{scarselli1998universal}%
  \BibitemOpen
  \bibfield  {author} {\bibinfo {author} {\bibfnamefont {F.}~\bibnamefont
  {Scarselli}}\ and\ \bibinfo {author} {\bibfnamefont {A.~C.}\ \bibnamefont
  {Tsoi}},\ }\href@noop {} {\bibfield  {journal} {\bibinfo  {journal} {Neural
  networks}\ }\textbf {\bibinfo {volume} {11}},\ \bibinfo {pages} {15}
  (\bibinfo {year} {1998})}\BibitemShut {NoStop}%
\bibitem [{\citenamefont {Casalicchio}\ \emph {et~al.}(2018)\citenamefont
  {Casalicchio}, \citenamefont {Molnar},\ and\ \citenamefont
  {Bischl}}]{casalicchio2018visualizing}%
  \BibitemOpen
  \bibfield  {author} {\bibinfo {author} {\bibfnamefont {G.}~\bibnamefont
  {Casalicchio}}, \bibinfo {author} {\bibfnamefont {C.}~\bibnamefont {Molnar}},
  \ and\ \bibinfo {author} {\bibfnamefont {B.}~\bibnamefont {Bischl}},\ }in\
  \href@noop {} {\emph {\bibinfo {booktitle} {Joint European Conference on
  Machine Learning and Knowledge Discovery in Databases}}}\ (\bibinfo
  {organization} {Springer},\ \bibinfo {year} {2018})\ pp.\ \bibinfo {pages}
  {655--670}\BibitemShut {NoStop}%
\bibitem [{\citenamefont {Wojtas}\ and\ \citenamefont
  {Chen}(2020)}]{wojtas2020feature}%
  \BibitemOpen
  \bibfield  {author} {\bibinfo {author} {\bibfnamefont {M.}~\bibnamefont
  {Wojtas}}\ and\ \bibinfo {author} {\bibfnamefont {K.}~\bibnamefont {Chen}},\
  }\href@noop {} {\bibfield  {journal} {\bibinfo  {journal} {Advances in Neural
  Information Processing Systems}\ }\textbf {\bibinfo {volume} {33}},\ \bibinfo
  {pages} {5105} (\bibinfo {year} {2020})}\BibitemShut {NoStop}%
\bibitem [{\citenamefont {Rengasamy}\ \emph {et~al.}(2021)\citenamefont
  {Rengasamy}, \citenamefont {Mase}, \citenamefont {Torres}, \citenamefont
  {Rothwell}, \citenamefont {Winkler},\ and\ \citenamefont
  {Figueredo}}]{rengasamy2021mechanistic}%
  \BibitemOpen
  \bibfield  {author} {\bibinfo {author} {\bibfnamefont {D.}~\bibnamefont
  {Rengasamy}}, \bibinfo {author} {\bibfnamefont {J.~M.}\ \bibnamefont {Mase}},
  \bibinfo {author} {\bibfnamefont {M.~T.}\ \bibnamefont {Torres}}, \bibinfo
  {author} {\bibfnamefont {B.}~\bibnamefont {Rothwell}}, \bibinfo {author}
  {\bibfnamefont {D.~A.}\ \bibnamefont {Winkler}}, \ and\ \bibinfo {author}
  {\bibfnamefont {G.~P.}\ \bibnamefont {Figueredo}},\ }\href@noop {} {\bibfield
   {journal} {\bibinfo  {journal} {arXiv preprint arXiv:2110.11713}\ }
  (\bibinfo {year} {2021})}\BibitemShut {NoStop}%
\bibitem [{\citenamefont {Cranmer}\ \emph
  {et~al.}(2020{\natexlab{a}})\citenamefont {Cranmer}, \citenamefont
  {Greydanus}, \citenamefont {Hoyer}, \citenamefont {Battaglia}, \citenamefont
  {Spergel},\ and\ \citenamefont {Ho}}]{cranmer2020lagrangian}%
  \BibitemOpen
  \bibfield  {author} {\bibinfo {author} {\bibfnamefont {M.}~\bibnamefont
  {Cranmer}}, \bibinfo {author} {\bibfnamefont {S.}~\bibnamefont {Greydanus}},
  \bibinfo {author} {\bibfnamefont {S.}~\bibnamefont {Hoyer}}, \bibinfo
  {author} {\bibfnamefont {P.}~\bibnamefont {Battaglia}}, \bibinfo {author}
  {\bibfnamefont {D.}~\bibnamefont {Spergel}}, \ and\ \bibinfo {author}
  {\bibfnamefont {S.}~\bibnamefont {Ho}},\ }\href@noop {} {\bibfield  {journal}
  {\bibinfo  {journal} {arXiv preprint arXiv:2003.04630}\ } (\bibinfo {year}
  {2020}{\natexlab{a}})}\BibitemShut {NoStop}%
\bibitem [{\citenamefont {Chen}\ \emph
  {et~al.}(2018{\natexlab{b}})\citenamefont {Chen}, \citenamefont {Rubanova},
  \citenamefont {Bettencourt},\ and\ \citenamefont
  {Duvenaud}}]{chen2018neural}%
  \BibitemOpen
  \bibfield  {author} {\bibinfo {author} {\bibfnamefont {R.~T.}\ \bibnamefont
  {Chen}}, \bibinfo {author} {\bibfnamefont {Y.}~\bibnamefont {Rubanova}},
  \bibinfo {author} {\bibfnamefont {J.}~\bibnamefont {Bettencourt}}, \ and\
  \bibinfo {author} {\bibfnamefont {D.~K.}\ \bibnamefont {Duvenaud}},\
  }\href@noop {} {\bibfield  {journal} {\bibinfo  {journal} {Advances in neural
  information processing systems}\ }\textbf {\bibinfo {volume} {31}} (\bibinfo
  {year} {2018}{\natexlab{b}})}\BibitemShut {NoStop}%
\bibitem [{\citenamefont {Liu}\ and\ \citenamefont
  {Tegmark}(2021)}]{liu2021machine}%
  \BibitemOpen
  \bibfield  {author} {\bibinfo {author} {\bibfnamefont {Z.}~\bibnamefont
  {Liu}}\ and\ \bibinfo {author} {\bibfnamefont {M.}~\bibnamefont {Tegmark}},\
  }\href@noop {} {\bibfield  {journal} {\bibinfo  {journal} {Physical Review
  Letters}\ }\textbf {\bibinfo {volume} {126}},\ \bibinfo {pages} {180604}
  (\bibinfo {year} {2021})}\BibitemShut {NoStop}%
\bibitem [{\citenamefont {Lee}\ and\ \citenamefont
  {Carlberg}(2019)}]{lee2019deep}%
  \BibitemOpen
  \bibfield  {author} {\bibinfo {author} {\bibfnamefont {K.}~\bibnamefont
  {Lee}}\ and\ \bibinfo {author} {\bibfnamefont {K.}~\bibnamefont {Carlberg}},\
  }\href@noop {} {\bibfield  {journal} {\bibinfo  {journal} {arXiv preprint
  arXiv:1909.09754}\ } (\bibinfo {year} {2019})}\BibitemShut {NoStop}%
\bibitem [{\citenamefont {Cranmer}\ \emph
  {et~al.}(2020{\natexlab{b}})\citenamefont {Cranmer}, \citenamefont
  {Sanchez~Gonzalez}, \citenamefont {Battaglia}, \citenamefont {Xu},
  \citenamefont {Cranmer}, \citenamefont {Spergel},\ and\ \citenamefont
  {Ho}}]{cranmer2020discovering}%
  \BibitemOpen
  \bibfield  {author} {\bibinfo {author} {\bibfnamefont {M.}~\bibnamefont
  {Cranmer}}, \bibinfo {author} {\bibfnamefont {A.}~\bibnamefont
  {Sanchez~Gonzalez}}, \bibinfo {author} {\bibfnamefont {P.}~\bibnamefont
  {Battaglia}}, \bibinfo {author} {\bibfnamefont {R.}~\bibnamefont {Xu}},
  \bibinfo {author} {\bibfnamefont {K.}~\bibnamefont {Cranmer}}, \bibinfo
  {author} {\bibfnamefont {D.}~\bibnamefont {Spergel}}, \ and\ \bibinfo
  {author} {\bibfnamefont {S.}~\bibnamefont {Ho}},\ }\href@noop {} {\bibfield
  {journal} {\bibinfo  {journal} {Advances in Neural Information Processing
  Systems}\ }\textbf {\bibinfo {volume} {33}},\ \bibinfo {pages} {17429}
  (\bibinfo {year} {2020}{\natexlab{b}})}\BibitemShut {NoStop}%
\bibitem [{\citenamefont {Udrescu}\ \emph {et~al.}(2020)\citenamefont
  {Udrescu}, \citenamefont {Tan}, \citenamefont {Feng}, \citenamefont {Neto},
  \citenamefont {Wu},\ and\ \citenamefont {Tegmark}}]{udrescu2020ai}%
  \BibitemOpen
  \bibfield  {author} {\bibinfo {author} {\bibfnamefont {S.-M.}\ \bibnamefont
  {Udrescu}}, \bibinfo {author} {\bibfnamefont {A.}~\bibnamefont {Tan}},
  \bibinfo {author} {\bibfnamefont {J.}~\bibnamefont {Feng}}, \bibinfo {author}
  {\bibfnamefont {O.}~\bibnamefont {Neto}}, \bibinfo {author} {\bibfnamefont
  {T.}~\bibnamefont {Wu}}, \ and\ \bibinfo {author} {\bibfnamefont
  {M.}~\bibnamefont {Tegmark}},\ }\href@noop {} {\bibfield  {journal} {\bibinfo
   {journal} {Advances in Neural Information Processing Systems}\ }\textbf
  {\bibinfo {volume} {33}},\ \bibinfo {pages} {4860} (\bibinfo {year}
  {2020})}\BibitemShut {NoStop}%
\bibitem [{\citenamefont {Bakarji}\ \emph {et~al.}(2022)\citenamefont
  {Bakarji}, \citenamefont {Champion}, \citenamefont {Kutz},\ and\
  \citenamefont {Brunton}}]{bakarji2022discovering}%
  \BibitemOpen
  \bibfield  {author} {\bibinfo {author} {\bibfnamefont {J.}~\bibnamefont
  {Bakarji}}, \bibinfo {author} {\bibfnamefont {K.}~\bibnamefont {Champion}},
  \bibinfo {author} {\bibfnamefont {J.~N.}\ \bibnamefont {Kutz}}, \ and\
  \bibinfo {author} {\bibfnamefont {S.~L.}\ \bibnamefont {Brunton}},\
  }\href@noop {} {\bibfield  {journal} {\bibinfo  {journal} {arXiv preprint
  arXiv:2201.05136}\ } (\bibinfo {year} {2022})}\BibitemShut {NoStop}%
\bibitem [{\citenamefont {Niu}\ \emph {et~al.}(2019)\citenamefont {Niu},
  \citenamefont {Liang}, \citenamefont {Sun}, \citenamefont {Long},
  \citenamefont {Niu} \emph {et~al.}}]{niu2019predictions}%
  \BibitemOpen
  \bibfield  {author} {\bibinfo {author} {\bibfnamefont {Z.}~\bibnamefont
  {Niu}}, \bibinfo {author} {\bibfnamefont {H.}~\bibnamefont {Liang}}, \bibinfo
  {author} {\bibfnamefont {B.}~\bibnamefont {Sun}}, \bibinfo {author}
  {\bibfnamefont {W.}~\bibnamefont {Long}}, \bibinfo {author} {\bibfnamefont
  {Y.}~\bibnamefont {Niu}},  \emph {et~al.},\ }\href@noop {} {\bibfield
  {journal} {\bibinfo  {journal} {Physical Review C}\ }\textbf {\bibinfo
  {volume} {99}},\ \bibinfo {pages} {064307} (\bibinfo {year}
  {2019})}\BibitemShut {NoStop}%
\bibitem [{\citenamefont {Bayram}\ \emph {et~al.}(2014)\citenamefont {Bayram},
  \citenamefont {Akkoyun},\ and\ \citenamefont {Kara}}]{bayram}%
  \BibitemOpen
  \bibfield  {author} {\bibinfo {author} {\bibfnamefont {T.}~\bibnamefont
  {Bayram}}, \bibinfo {author} {\bibfnamefont {S.}~\bibnamefont {Akkoyun}}, \
  and\ \bibinfo {author} {\bibfnamefont {S.~O.}\ \bibnamefont {Kara}},\ }\href
  {\doibase 10.1016/j.anucene.2013.07.039} {\bibfield  {journal} {\bibinfo
  {journal} {Annals of Nuclear Energy}\ }\textbf {\bibinfo {volume} {63}},\
  \bibinfo {pages} {172} (\bibinfo {year} {2014})},\ \Eprint
  {http://arxiv.org/abs/1301.2407} {arXiv:1301.2407 [nucl-th]} \BibitemShut
  {NoStop}%
\bibitem [{\citenamefont {Cruz}(2019)}]{cruz2019clustering}%
  \BibitemOpen
  \bibfield  {author} {\bibinfo {author} {\bibfnamefont {M.}~\bibnamefont
  {Cruz}},\ }\href@noop {} {\bibfield  {journal} {\bibinfo  {journal} {Bulletin
  of the American Physical Society}\ }\textbf {\bibinfo {volume} {64}}
  (\bibinfo {year} {2019})}\BibitemShut {NoStop}%
\bibitem [{\citenamefont {Costiris}\ \emph {et~al.}(2007)\citenamefont
  {Costiris}, \citenamefont {Mavrommatis}, \citenamefont {Gernoth},\ and\
  \citenamefont {Clark}}]{costiris2007global}%
  \BibitemOpen
  \bibfield  {author} {\bibinfo {author} {\bibfnamefont {N.}~\bibnamefont
  {Costiris}}, \bibinfo {author} {\bibfnamefont {E.}~\bibnamefont
  {Mavrommatis}}, \bibinfo {author} {\bibfnamefont {K.}~\bibnamefont
  {Gernoth}}, \ and\ \bibinfo {author} {\bibfnamefont {J.}~\bibnamefont
  {Clark}},\ }\href@noop {} {\bibfield  {journal} {\bibinfo  {journal} {arXiv
  preprint nucl-th/0701096}\ } (\bibinfo {year} {2007})}\BibitemShut {NoStop}%
\bibitem [{\citenamefont {Gao}\ \emph {et~al.}(2021)\citenamefont {Gao},
  \citenamefont {Wang}, \citenamefont {L{\"u}}, \citenamefont {Li},
  \citenamefont {Shen},\ and\ \citenamefont {Liu}}]{gao2021machine}%
  \BibitemOpen
  \bibfield  {author} {\bibinfo {author} {\bibfnamefont {Z.-P.}\ \bibnamefont
  {Gao}}, \bibinfo {author} {\bibfnamefont {Y.-J.}\ \bibnamefont {Wang}},
  \bibinfo {author} {\bibfnamefont {H.-L.}\ \bibnamefont {L{\"u}}}, \bibinfo
  {author} {\bibfnamefont {Q.-F.}\ \bibnamefont {Li}}, \bibinfo {author}
  {\bibfnamefont {C.-W.}\ \bibnamefont {Shen}}, \ and\ \bibinfo {author}
  {\bibfnamefont {L.}~\bibnamefont {Liu}},\ }\href@noop {} {\bibfield
  {journal} {\bibinfo  {journal} {Nuclear Science and Techniques}\ }\textbf
  {\bibinfo {volume} {32}},\ \bibinfo {pages} {1} (\bibinfo {year}
  {2021})}\BibitemShut {NoStop}%
\bibitem [{\citenamefont {Athanassopoulos}\ \emph {et~al.}(2005)\citenamefont
  {Athanassopoulos}, \citenamefont {Mavrommatis}, \citenamefont {Gernoth},\
  and\ \citenamefont {Clark}}]{athanosso-1}%
  \BibitemOpen
  \bibfield  {author} {\bibinfo {author} {\bibfnamefont {S.}~\bibnamefont
  {Athanassopoulos}}, \bibinfo {author} {\bibfnamefont {E.}~\bibnamefont
  {Mavrommatis}}, \bibinfo {author} {\bibfnamefont {K.~A.}\ \bibnamefont
  {Gernoth}}, \ and\ \bibinfo {author} {\bibfnamefont {J.~W.}\ \bibnamefont
  {Clark}},\ }\href {\doibase 10.48550/ARXIV.NUCL-TH/0509075} {\enquote
  {\bibinfo {title} {One and two proton separation energies from nuclear mass
  systematics using neural networks},}\ } (\bibinfo {year} {2005})\BibitemShut
  {NoStop}%
\bibitem [{\citenamefont {Athanassopoulos}\ \emph {et~al.}(2004)\citenamefont
  {Athanassopoulos}, \citenamefont {Mavrommatis}, \citenamefont {Gernoth},\
  and\ \citenamefont {Clark}}]{athanosso-2}%
  \BibitemOpen
  \bibfield  {author} {\bibinfo {author} {\bibfnamefont {S.}~\bibnamefont
  {Athanassopoulos}}, \bibinfo {author} {\bibfnamefont {E.}~\bibnamefont
  {Mavrommatis}}, \bibinfo {author} {\bibfnamefont {K.}~\bibnamefont
  {Gernoth}}, \ and\ \bibinfo {author} {\bibfnamefont {J.}~\bibnamefont
  {Clark}},\ }\href {\doibase 10.1016/j.nuclphysa.2004.08.006} {\bibfield
  {journal} {\bibinfo  {journal} {Nuclear Physics A}\ }\textbf {\bibinfo
  {volume} {743}},\ \bibinfo {pages} {222} (\bibinfo {year}
  {2004})}\BibitemShut {NoStop}%
\bibitem [{\citenamefont {Bass}\ \emph {et~al.}(1996)\citenamefont {Bass},
  \citenamefont {Bischoff}, \citenamefont {Maruhn}, \citenamefont {Stoecker},\
  and\ \citenamefont {Greiner}}]{Bass:1996ez}%
  \BibitemOpen
  \bibfield  {author} {\bibinfo {author} {\bibfnamefont {S.~A.}\ \bibnamefont
  {Bass}}, \bibinfo {author} {\bibfnamefont {A.}~\bibnamefont {Bischoff}},
  \bibinfo {author} {\bibfnamefont {J.~A.}\ \bibnamefont {Maruhn}}, \bibinfo
  {author} {\bibfnamefont {H.}~\bibnamefont {Stoecker}}, \ and\ \bibinfo
  {author} {\bibfnamefont {W.}~\bibnamefont {Greiner}},\ }\href {\doibase
  10.1103/PhysRevC.53.2358} {\bibfield  {journal} {\bibinfo  {journal} {Phys.
  Rev. C}\ }\textbf {\bibinfo {volume} {53}},\ \bibinfo {pages} {2358}
  (\bibinfo {year} {1996})},\ \Eprint {http://arxiv.org/abs/nucl-th/9601024}
  {arXiv:nucl-th/9601024} \BibitemShut {NoStop}%
\bibitem [{\citenamefont {Akkoyun}(2020)}]{nimb}%
  \BibitemOpen
  \bibfield  {author} {\bibinfo {author} {\bibfnamefont {S.}~\bibnamefont
  {Akkoyun}},\ }\href {\doibase 10.1016/j.nimb.2019.11.014} {\bibfield
  {journal} {\bibinfo  {journal} {Nucl. Instrum. Meth. B}\ }\textbf {\bibinfo
  {volume} {462}},\ \bibinfo {pages} {51} (\bibinfo {year} {2020})},\ \Eprint
  {http://arxiv.org/abs/1907.00579} {arXiv:1907.00579 [nucl-ex]} \BibitemShut
  {NoStop}%
\bibitem [{\citenamefont {Akkoyun}\ \emph {et~al.}(2013)\citenamefont
  {Akkoyun}, \citenamefont {Bayram}, \citenamefont {Kara},\ and\ \citenamefont
  {Sinan}}]{radii-1}%
  \BibitemOpen
  \bibfield  {author} {\bibinfo {author} {\bibfnamefont {S.}~\bibnamefont
  {Akkoyun}}, \bibinfo {author} {\bibfnamefont {T.}~\bibnamefont {Bayram}},
  \bibinfo {author} {\bibfnamefont {S.~O.}\ \bibnamefont {Kara}}, \ and\
  \bibinfo {author} {\bibfnamefont {A.}~\bibnamefont {Sinan}},\ }\href
  {\doibase 10.1088/0954-3899/40/5/055106} {\bibfield  {journal} {\bibinfo
  {journal} {J. Phys. G}\ }\textbf {\bibinfo {volume} {40}},\ \bibinfo {pages}
  {055106} (\bibinfo {year} {2013})},\ \Eprint {http://arxiv.org/abs/1212.6319}
  {arXiv:1212.6319 [nucl-th]} \BibitemShut {NoStop}%
\bibitem [{\citenamefont {Costiris}\ \emph {et~al.}(2009)\citenamefont
  {Costiris}, \citenamefont {Mavrommatis}, \citenamefont {Gernoth},\ and\
  \citenamefont {Clark}}]{cost}%
  \BibitemOpen
  \bibfield  {author} {\bibinfo {author} {\bibfnamefont {N.~J.}\ \bibnamefont
  {Costiris}}, \bibinfo {author} {\bibfnamefont {E.}~\bibnamefont
  {Mavrommatis}}, \bibinfo {author} {\bibfnamefont {K.~A.}\ \bibnamefont
  {Gernoth}}, \ and\ \bibinfo {author} {\bibfnamefont {J.~W.}\ \bibnamefont
  {Clark}},\ }\href {\doibase 10.1103/PhysRevC.80.044332} {\bibfield  {journal}
  {\bibinfo  {journal} {Phys. Rev. C}\ }\textbf {\bibinfo {volume} {80}},\
  \bibinfo {pages} {044332} (\bibinfo {year} {2009})},\ \Eprint
  {http://arxiv.org/abs/0806.2850} {arXiv:0806.2850 [nucl-th]} \BibitemShut
  {NoStop}%
\bibitem [{\citenamefont {Akkoyun}\ and\ \citenamefont
  {Bayram}(2014)}]{barrier}%
  \BibitemOpen
  \bibfield  {author} {\bibinfo {author} {\bibfnamefont {S.}~\bibnamefont
  {Akkoyun}}\ and\ \bibinfo {author} {\bibfnamefont {T.}~\bibnamefont
  {Bayram}},\ }\href@noop {} {\bibfield  {journal} {\bibinfo  {journal}
  {International Journal of Modern Physics E}\ }\textbf {\bibinfo {volume}
  {23}},\ \bibinfo {pages} {1450064} (\bibinfo {year} {2014})}\BibitemShut
  {NoStop}%
\bibitem [{\citenamefont {Akkoyun}\ \emph {et~al.}(2014)\citenamefont
  {Akkoyun}, \citenamefont {Bayram},\ and\ \citenamefont {Turker}}]{beta}%
  \BibitemOpen
  \bibfield  {author} {\bibinfo {author} {\bibfnamefont {S.}~\bibnamefont
  {Akkoyun}}, \bibinfo {author} {\bibfnamefont {T.}~\bibnamefont {Bayram}}, \
  and\ \bibinfo {author} {\bibfnamefont {T.}~\bibnamefont {Turker}},\
  }\href@noop {} {\bibfield  {journal} {\bibinfo  {journal} {radiation Physics
  and Chemistry}\ }\textbf {\bibinfo {volume} {96}},\ \bibinfo {pages} {186}
  (\bibinfo {year} {2014})}\BibitemShut {NoStop}%
\bibitem [{\citenamefont {Tondeur}\ \emph {et~al.}(2000)\citenamefont
  {Tondeur}, \citenamefont {Goriely}, \citenamefont {Pearson},\ and\
  \citenamefont {Onsi}}]{Tondeur}%
  \BibitemOpen
  \bibfield  {author} {\bibinfo {author} {\bibfnamefont {F.}~\bibnamefont
  {Tondeur}}, \bibinfo {author} {\bibfnamefont {S.}~\bibnamefont {Goriely}},
  \bibinfo {author} {\bibfnamefont {J.~M.}\ \bibnamefont {Pearson}}, \ and\
  \bibinfo {author} {\bibfnamefont {M.}~\bibnamefont {Onsi}},\ }\href {\doibase
  10.1103/PhysRevC.62.024308} {\bibfield  {journal} {\bibinfo  {journal} {Phys.
  Rev. C}\ }\textbf {\bibinfo {volume} {62}},\ \bibinfo {pages} {024308}
  (\bibinfo {year} {2000})}\BibitemShut {NoStop}%
\bibitem [{\citenamefont {Wang}(2021)}]{AME2020}%
  \BibitemOpen
  \bibfield  {author} {\bibinfo {author} {\bibfnamefont {M.}~\bibnamefont
  {Wang}},\ }\href {\doibase 10.1088/1674-1137/abddaf} {\bibfield  {journal}
  {\bibinfo  {journal} {Chinese Physics C}\ }\textbf {\bibinfo {volume} {45}}
  (\bibinfo {year} {2021}),\ 10.1088/1674-1137/abddaf}\BibitemShut {NoStop}%
\bibitem [{\citenamefont {Garg}\ and\ \citenamefont
  {Tai}(2012)}]{garg2012comparison}%
  \BibitemOpen
  \bibfield  {author} {\bibinfo {author} {\bibfnamefont {A.}~\bibnamefont
  {Garg}}\ and\ \bibinfo {author} {\bibfnamefont {K.}~\bibnamefont {Tai}},\
  }in\ \href@noop {} {\emph {\bibinfo {booktitle} {2012 Proceedings of
  International Conference on Modelling, Identification and Control}}}\
  (\bibinfo {organization} {IEEE},\ \bibinfo {year} {2012})\ pp.\ \bibinfo
  {pages} {353--358}\BibitemShut {NoStop}%
\bibitem [{\citenamefont {Jolliffe}(1982)}]{jolliffe1982note}%
  \BibitemOpen
  \bibfield  {author} {\bibinfo {author} {\bibfnamefont {I.~T.}\ \bibnamefont
  {Jolliffe}},\ }\href@noop {} {\bibfield  {journal} {\bibinfo  {journal}
  {Journal of the Royal Statistical Society: Series C (Applied Statistics)}\
  }\textbf {\bibinfo {volume} {31}},\ \bibinfo {pages} {300} (\bibinfo {year}
  {1982})}\BibitemShut {NoStop}%
\bibitem [{\citenamefont {Akiba}\ \emph
  {et~al.}(2019{\natexlab{a}})\citenamefont {Akiba}, \citenamefont {Sano},
  \citenamefont {Yanase}, \citenamefont {Ohta},\ and\ \citenamefont
  {Koyama}}]{DBLP:journals/corr/abs-1907-10902}%
  \BibitemOpen
  \bibfield  {author} {\bibinfo {author} {\bibfnamefont {T.}~\bibnamefont
  {Akiba}}, \bibinfo {author} {\bibfnamefont {S.}~\bibnamefont {Sano}},
  \bibinfo {author} {\bibfnamefont {T.}~\bibnamefont {Yanase}}, \bibinfo
  {author} {\bibfnamefont {T.}~\bibnamefont {Ohta}}, \ and\ \bibinfo {author}
  {\bibfnamefont {M.}~\bibnamefont {Koyama}},\ }\href
  {http://arxiv.org/abs/1907.10902} {\bibfield  {journal} {\bibinfo  {journal}
  {CoRR}\ }\textbf {\bibinfo {volume} {abs/1907.10902}} (\bibinfo {year}
  {2019}{\natexlab{a}})},\ \Eprint {http://arxiv.org/abs/1907.10902}
  {1907.10902} \BibitemShut {NoStop}%
\bibitem [{\citenamefont {Nguyen}\ \emph {et~al.}(2021)\citenamefont {Nguyen},
  \citenamefont {Pham}, \citenamefont {Ngo}, \citenamefont {Ngo},\ and\
  \citenamefont {Pham}}]{nguyen2021analysis}%
  \BibitemOpen
  \bibfield  {author} {\bibinfo {author} {\bibfnamefont {A.}~\bibnamefont
  {Nguyen}}, \bibinfo {author} {\bibfnamefont {K.}~\bibnamefont {Pham}},
  \bibinfo {author} {\bibfnamefont {D.}~\bibnamefont {Ngo}}, \bibinfo {author}
  {\bibfnamefont {T.}~\bibnamefont {Ngo}}, \ and\ \bibinfo {author}
  {\bibfnamefont {L.}~\bibnamefont {Pham}},\ }in\ \href@noop {} {\emph
  {\bibinfo {booktitle} {2021 International Conference on System Science and
  Engineering (ICSSE)}}}\ (\bibinfo {organization} {IEEE},\ \bibinfo {year}
  {2021})\ pp.\ \bibinfo {pages} {215--220}\BibitemShut {NoStop}%
\bibitem [{\citenamefont {Garbin}\ \emph {et~al.}(2020)\citenamefont {Garbin},
  \citenamefont {Zhu},\ and\ \citenamefont {Marques}}]{garbin2020dropout}%
  \BibitemOpen
  \bibfield  {author} {\bibinfo {author} {\bibfnamefont {C.}~\bibnamefont
  {Garbin}}, \bibinfo {author} {\bibfnamefont {X.}~\bibnamefont {Zhu}}, \ and\
  \bibinfo {author} {\bibfnamefont {O.}~\bibnamefont {Marques}},\ }\href@noop
  {} {\bibfield  {journal} {\bibinfo  {journal} {Multimedia Tools and
  Applications}\ }\textbf {\bibinfo {volume} {79}} (\bibinfo {year}
  {2020})}\BibitemShut {NoStop}%
\bibitem [{\citenamefont {Akiba}\ \emph
  {et~al.}(2019{\natexlab{b}})\citenamefont {Akiba}, \citenamefont {Sano},
  \citenamefont {Yanase}, \citenamefont {Ohta},\ and\ \citenamefont
  {Koyama}}]{akiba2019optuna}%
  \BibitemOpen
  \bibfield  {author} {\bibinfo {author} {\bibfnamefont {T.}~\bibnamefont
  {Akiba}}, \bibinfo {author} {\bibfnamefont {S.}~\bibnamefont {Sano}},
  \bibinfo {author} {\bibfnamefont {T.}~\bibnamefont {Yanase}}, \bibinfo
  {author} {\bibfnamefont {T.}~\bibnamefont {Ohta}}, \ and\ \bibinfo {author}
  {\bibfnamefont {M.}~\bibnamefont {Koyama}},\ }in\ \href@noop {} {\emph
  {\bibinfo {booktitle} {Proceedings of the 25th ACM SIGKDD international
  conference on knowledge discovery \& data mining}}}\ (\bibinfo {year}
  {2019})\ pp.\ \bibinfo {pages} {2623--2631}\BibitemShut {NoStop}%
\bibitem [{\citenamefont {Smith}(2017)}]{smith2017cyclical}%
  \BibitemOpen
  \bibfield  {author} {\bibinfo {author} {\bibfnamefont {L.~N.}\ \bibnamefont
  {Smith}},\ }in\ \href@noop {} {\emph {\bibinfo {booktitle} {2017 IEEE winter
  conference on applications of computer vision (WACV)}}}\ (\bibinfo
  {organization} {IEEE},\ \bibinfo {year} {2017})\ pp.\ \bibinfo {pages}
  {464--472}\BibitemShut {NoStop}%
\bibitem [{\citenamefont {Falcon}\ and\ \citenamefont {{The PyTorch Lightning
  team}}(2019)}]{Falcon_PyTorch_Lightning_2019}%
  \BibitemOpen
  \bibfield  {author} {\bibinfo {author} {\bibfnamefont {W.}~\bibnamefont
  {Falcon}}\ and\ \bibinfo {author} {\bibnamefont {{The PyTorch Lightning
  team}}},\ }\href {\doibase 10.5281/zenodo.3828935} {\enquote {\bibinfo
  {title} {{PyTorch Lightning}},}\ } (\bibinfo {year} {2019})\BibitemShut
  {NoStop}%
\bibitem [{\citenamefont {Rajendra}\ \emph {et~al.}(2021)\citenamefont
  {Rajendra}, \citenamefont {Ravi.~PVN},\ and\ \citenamefont
  {Naidu~T}}]{rajendra2021optimization}%
  \BibitemOpen
  \bibfield  {author} {\bibinfo {author} {\bibfnamefont {P.}~\bibnamefont
  {Rajendra}}, \bibinfo {author} {\bibfnamefont {H.}~\bibnamefont {Ravi.~PVN}},
  \ and\ \bibinfo {author} {\bibfnamefont {G.}~\bibnamefont {Naidu~T}},\ }in\
  \href@noop {} {\emph {\bibinfo {booktitle} {AIP Conference Proceedings}}},\
  Vol.\ \bibinfo {volume} {2375}\ (\bibinfo {organization} {AIP Publishing
  LLC},\ \bibinfo {year} {2021})\ p.\ \bibinfo {pages} {020034}\BibitemShut
  {NoStop}%
\bibitem [{\citenamefont {Chen}\ and\ \citenamefont
  {Guestrin}(2016)}]{Chen_2016}%
  \BibitemOpen
  \bibfield  {author} {\bibinfo {author} {\bibfnamefont {T.}~\bibnamefont
  {Chen}}\ and\ \bibinfo {author} {\bibfnamefont {C.}~\bibnamefont
  {Guestrin}},\ }in\ \href {\doibase 10.1145/2939672.2939785} {\emph {\bibinfo
  {booktitle} {Proceedings of the 22nd {ACM} {SIGKDD} International Conference
  on Knowledge Discovery and Data Mining}}}\ (\bibinfo  {publisher} {{ACM}},\
  \bibinfo {year} {2016})\BibitemShut {NoStop}%
\bibitem [{\citenamefont {Shwartz-Ziv}\ and\ \citenamefont
  {Armon}(2021)}]{dl_not_all}%
  \BibitemOpen
  \bibfield  {author} {\bibinfo {author} {\bibfnamefont {R.}~\bibnamefont
  {Shwartz-Ziv}}\ and\ \bibinfo {author} {\bibfnamefont {A.}~\bibnamefont
  {Armon}},\ }\href {\doibase 10.48550/ARXIV.2106.03253} {\enquote {\bibinfo
  {title} {Tabular data: Deep learning is not all you need},}\ } (\bibinfo
  {year} {2021})\BibitemShut {NoStop}%
\bibitem [{\citenamefont {Akiba}\ \emph
  {et~al.}(2019{\natexlab{c}})\citenamefont {Akiba}, \citenamefont {Sano},
  \citenamefont {Yanase}, \citenamefont {Ohta},\ and\ \citenamefont
  {Koyama}}]{optuna}%
  \BibitemOpen
  \bibfield  {author} {\bibinfo {author} {\bibfnamefont {T.}~\bibnamefont
  {Akiba}}, \bibinfo {author} {\bibfnamefont {S.}~\bibnamefont {Sano}},
  \bibinfo {author} {\bibfnamefont {T.}~\bibnamefont {Yanase}}, \bibinfo
  {author} {\bibfnamefont {T.}~\bibnamefont {Ohta}}, \ and\ \bibinfo {author}
  {\bibfnamefont {M.}~\bibnamefont {Koyama}},\ }\href {\doibase
  10.48550/ARXIV.1907.10902} {\enquote {\bibinfo {title} {Optuna: A
  next-generation hyperparameter optimization framework},}\ } (\bibinfo {year}
  {2019}{\natexlab{c}})\BibitemShut {NoStop}%
\bibitem [{\citenamefont {Vaswani}\ \emph {et~al.}(2017)\citenamefont
  {Vaswani}, \citenamefont {Shazeer}, \citenamefont {Parmar}, \citenamefont
  {Uszkoreit}, \citenamefont {Jones}, \citenamefont {Gomez}, \citenamefont
  {Kaiser},\ and\ \citenamefont {Polosukhin}}]{vaswani2017attention}%
  \BibitemOpen
  \bibfield  {author} {\bibinfo {author} {\bibfnamefont {A.}~\bibnamefont
  {Vaswani}}, \bibinfo {author} {\bibfnamefont {N.}~\bibnamefont {Shazeer}},
  \bibinfo {author} {\bibfnamefont {N.}~\bibnamefont {Parmar}}, \bibinfo
  {author} {\bibfnamefont {J.}~\bibnamefont {Uszkoreit}}, \bibinfo {author}
  {\bibfnamefont {L.}~\bibnamefont {Jones}}, \bibinfo {author} {\bibfnamefont
  {A.~N.}\ \bibnamefont {Gomez}}, \bibinfo {author} {\bibfnamefont
  {{\L}.}~\bibnamefont {Kaiser}}, \ and\ \bibinfo {author} {\bibfnamefont
  {I.}~\bibnamefont {Polosukhin}},\ }\href@noop {} {\bibfield  {journal}
  {\bibinfo  {journal} {Advances in neural information processing systems}\
  }\textbf {\bibinfo {volume} {30}} (\bibinfo {year} {2017})}\BibitemShut
  {NoStop}%
\bibitem [{\citenamefont {Arik}\ and\ \citenamefont {Pfister}(2019)}]{tabnet}%
  \BibitemOpen
  \bibfield  {author} {\bibinfo {author} {\bibfnamefont {S.~O.}\ \bibnamefont
  {Arik}}\ and\ \bibinfo {author} {\bibfnamefont {T.}~\bibnamefont {Pfister}},\
  }\href {\doibase 10.48550/ARXIV.1908.07442} {\enquote {\bibinfo {title}
  {Tabnet: Attentive interpretable tabular learning},}\ } (\bibinfo {year}
  {2019})\BibitemShut {NoStop}%
\bibitem [{\citenamefont {Kim}\ \emph {et~al.}(2020)\citenamefont {Kim},
  \citenamefont {Kim},\ and\ \citenamefont {Kim}}]{test_time_aug}%
  \BibitemOpen
  \bibfield  {author} {\bibinfo {author} {\bibfnamefont {I.}~\bibnamefont
  {Kim}}, \bibinfo {author} {\bibfnamefont {Y.}~\bibnamefont {Kim}}, \ and\
  \bibinfo {author} {\bibfnamefont {S.}~\bibnamefont {Kim}},\ }\href {\doibase
  10.48550/ARXIV.2010.11422} {\enquote {\bibinfo {title} {Learning loss for
  test-time augmentation},}\ } (\bibinfo {year} {2020})\BibitemShut {NoStop}%
\bibitem [{\citenamefont {Lundberg}\ and\ \citenamefont
  {Lee}(2017)}]{NIPS2017_7062}%
  \BibitemOpen
  \bibfield  {author} {\bibinfo {author} {\bibfnamefont {S.~M.}\ \bibnamefont
  {Lundberg}}\ and\ \bibinfo {author} {\bibfnamefont {S.-I.}\ \bibnamefont
  {Lee}},\ }in\ \href
  {http://papers.nips.cc/paper/7062-a-unified-approach-to-interpreting-model-predictions.pdf}
  {\emph {\bibinfo {booktitle} {Advances in Neural Information Processing
  Systems 30}}},\ \bibinfo {editor} {edited by\ \bibinfo {editor}
  {\bibfnamefont {I.}~\bibnamefont {Guyon}}, \bibinfo {editor} {\bibfnamefont
  {U.~V.}\ \bibnamefont {Luxburg}}, \bibinfo {editor} {\bibfnamefont
  {S.}~\bibnamefont {Bengio}}, \bibinfo {editor} {\bibfnamefont
  {H.}~\bibnamefont {Wallach}}, \bibinfo {editor} {\bibfnamefont
  {R.}~\bibnamefont {Fergus}}, \bibinfo {editor} {\bibfnamefont
  {S.}~\bibnamefont {Vishwanathan}}, \ and\ \bibinfo {editor} {\bibfnamefont
  {R.}~\bibnamefont {Garnett}}}\ (\bibinfo  {publisher} {Curran Associates,
  Inc.},\ \bibinfo {year} {2017})\ pp.\ \bibinfo {pages}
  {4765--4774}\BibitemShut {NoStop}%
\bibitem [{\citenamefont {Lundberg}\ \emph {et~al.}(2020)\citenamefont
  {Lundberg}, \citenamefont {Erion}, \citenamefont {Chen}, \citenamefont
  {DeGrave}, \citenamefont {Prutkin}, \citenamefont {Nair}, \citenamefont
  {Katz}, \citenamefont {Himmelfarb}, \citenamefont {Bansal},\ and\
  \citenamefont {Lee}}]{lundberg2020local2global}%
  \BibitemOpen
  \bibfield  {author} {\bibinfo {author} {\bibfnamefont {S.~M.}\ \bibnamefont
  {Lundberg}}, \bibinfo {author} {\bibfnamefont {G.}~\bibnamefont {Erion}},
  \bibinfo {author} {\bibfnamefont {H.}~\bibnamefont {Chen}}, \bibinfo {author}
  {\bibfnamefont {A.}~\bibnamefont {DeGrave}}, \bibinfo {author} {\bibfnamefont
  {J.~M.}\ \bibnamefont {Prutkin}}, \bibinfo {author} {\bibfnamefont
  {B.}~\bibnamefont {Nair}}, \bibinfo {author} {\bibfnamefont {R.}~\bibnamefont
  {Katz}}, \bibinfo {author} {\bibfnamefont {J.}~\bibnamefont {Himmelfarb}},
  \bibinfo {author} {\bibfnamefont {N.}~\bibnamefont {Bansal}}, \ and\ \bibinfo
  {author} {\bibfnamefont {S.-I.}\ \bibnamefont {Lee}},\ }\href@noop {}
  {\bibfield  {journal} {\bibinfo  {journal} {Nature Machine Intelligence}\
  }\textbf {\bibinfo {volume} {2}},\ \bibinfo {pages} {2522} (\bibinfo {year}
  {2020})}\BibitemShut {NoStop}%
\bibitem [{\citenamefont {Brown}\ and\ \citenamefont
  {Wildenthal}(1988)}]{brown1988status}%
  \BibitemOpen
  \bibfield  {author} {\bibinfo {author} {\bibfnamefont {B.~A.}\ \bibnamefont
  {Brown}}\ and\ \bibinfo {author} {\bibfnamefont {B.}~\bibnamefont
  {Wildenthal}},\ }\href@noop {} {\bibfield  {journal} {\bibinfo  {journal}
  {Annual Review of Nuclear and Particle Science}\ }\textbf {\bibinfo {volume}
  {38}},\ \bibinfo {pages} {29} (\bibinfo {year} {1988})}\BibitemShut {NoStop}%
\bibitem [{\citenamefont {Lilley}(2013)}]{lilley2013nuclear}%
  \BibitemOpen
  \bibfield  {author} {\bibinfo {author} {\bibfnamefont {J.}~\bibnamefont
  {Lilley}},\ }\href@noop {} {\emph {\bibinfo {title} {Nuclear physics:
  principles and applications}}}\ (\bibinfo  {publisher} {John Wiley \& Sons},\
  \bibinfo {year} {2013})\BibitemShut {NoStop}%
\end{thebibliography}%

\appendix
\section{Data distribution}\label{sec:datadist}
\begin{figure*}[!h]
   \centering
   \includegraphics[scale=0.35]{
   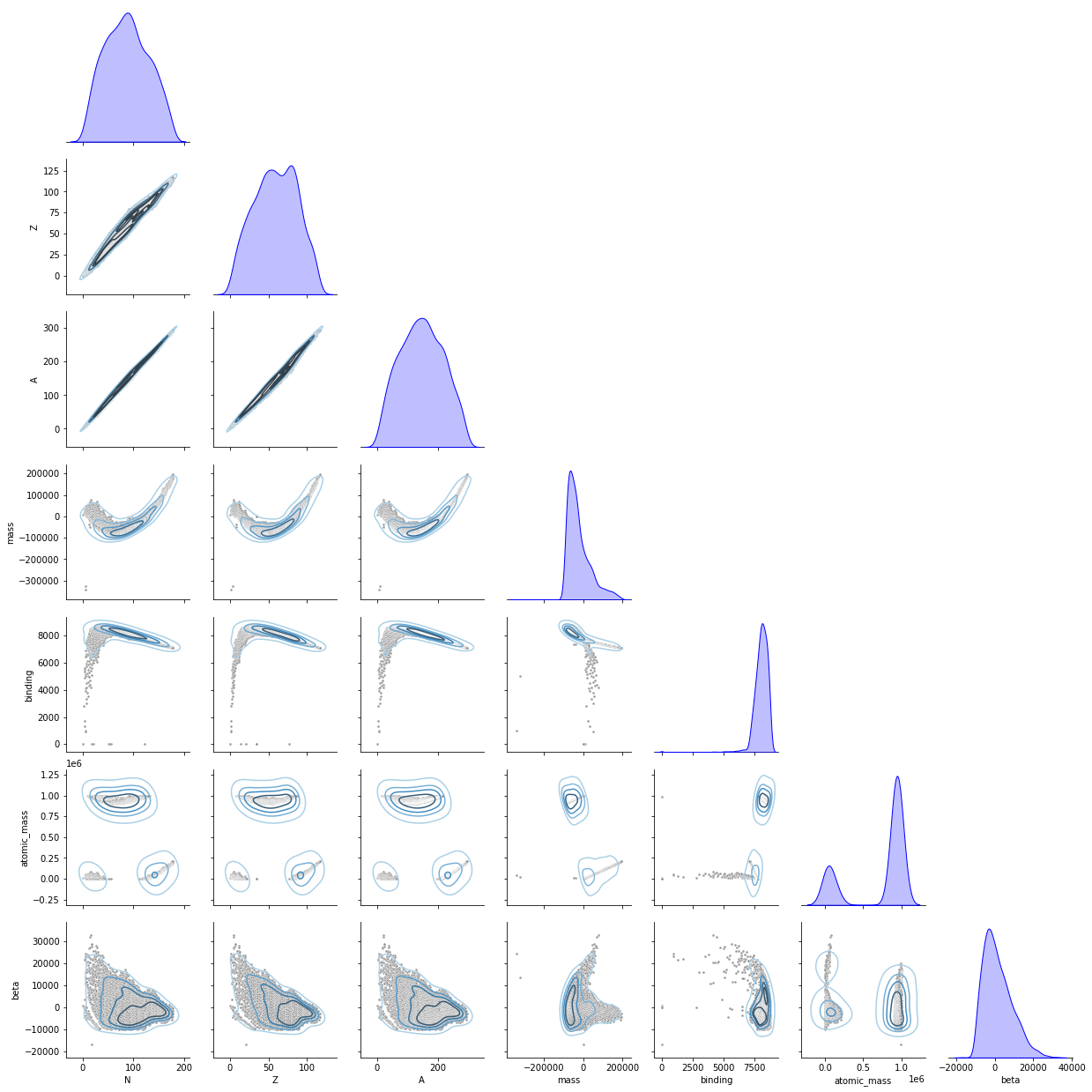}
   \caption{\centering Initial distribution of the atomic features in 
   the AME2020 Dataset~\cite{AME2020}.}
   \label{fig:distribution}
\end{figure*}

The tabular data used in this work compiles the atomic features published 
in \cite{AME2020}. 
It contains the features:

\begin{itemize}
   \item Neutron number $N$: Number of neutrons in each isotope.
   \item Proton number $Z$: Number of protons in each isotope.
   \item Mass number $A$: Atomic mass (N+Z).
   \item Mass excess $M_e$: Difference between measured mass and the mass 
   number ($A$).
   \item Binding Energy $B_E$: Measured per nucleon, refers to the
   ionization potential.
\end{itemize}
The distribution of the features is shown in Fig.~\ref{fig:distribution},
where no clear relationship between $Q_\beta$ is evident, but 
well-distributed samples for the atomic features are evident, both in
scattered and in the histograms.

\end{document}